\documentclass[
aps,
reprint,
superscriptaddress,
amsmath,
amssymb,
prapplied,
floatfix,
]{revtex4-2}
\usepackage{bm}
\usepackage{graphicx}
\usepackage{latexsym}
\usepackage{array}
\usepackage{amsfonts}
\usepackage{mathtools}  
\usepackage{makecell}   
\usepackage{hhline}     
\usepackage{amsthm}
\usepackage{amscd}

\usepackage{xcolor}

\usepackage{verbatim}
\usepackage{hyperref}
\usepackage[normalem]{ulem}
\graphicspath{{images/}}

 \newcommand{\angstrom}{\mbox{\normalfont\AA}}

\usepackage{etoolbox}
\apptocmd{\sloppy}{\hbadness 10000\relax}{}{}

\begin{document}

\preprint{APS/123-QED}

\title{Hysteresis and chaos in anomalous Josephson junctions without capacitance}
\author{A. A. Mazanik}
\affiliation{Donostia International Physics Center (DIPC), 20018 Donostia–San Sebasti\'an, Spain}
\affiliation{Bogoliubov Laboratory of Theoretical Physics, Joint Institute for Nuclear Research, \\ 141980, Dubna, Russian Federation}
\author{A. E. Botha}
\affiliation{Department of Physics, Unisa Science Campus, University of South Africa, Private Bag X6, Roodepoort, Johannesburg 1710, South Africa}
\author{I. R. Rahmonov}
\affiliation{Bogoliubov Laboratory of Theoretical Physics, Joint Institute for Nuclear Research, \\ 141980, Dubna, Russian Federation}
\affiliation{Dubna State University, 141980, Dubna,  Russian Federation}
\affiliation{Moscow Institute of Physics and Technology, 141700, \\ Dolgoprudny, Russian Federation}
\author{Yu. M. Shukrinov}
\affiliation{Bogoliubov Laboratory of Theoretical Physics, Joint Institute for Nuclear Research, \\ 141980, Dubna, Russian Federation}
\affiliation{Dubna State University, 141980, Dubna,  Russian Federation}
\affiliation{Moscow Institute of Physics and Technology, 141700, \\ Dolgoprudny, Russian Federation}

\date{\today}
\begin{abstract}
Usually, overdamped Josephson junctions do not exhibit chaotic behavior in their phase dynamics, either because the phase space dimension is less than three (as in the case of a single overdamped ac-driven junction) or due
to the general tendency of systems to become less chaotic with increasing dissipation (as in the case of coupled overdamped junctions). Here we consider the so-called $\varphi_0$ superconductor/ferromagnet/superconductor Josephson junction in which the current flowing through the junction may induce magnetization dynamics in the ferromagnetic interlayer. We find that due to the induced magnetization dynamics, even in the overdamped limit, i.e. for a junction without capacitance, the junction may exhibit chaos and hysteresis that in some cases leads to multiple branches in its current-voltage characteristics. We also show that pulsed current signals can be used to switch between the different voltage states, even in the presence of added thermal noise. Such switching could be used in cryogenic memory components.
\end{abstract}

\maketitle

\section{Introduction} \label{sec:Introduction}
As is well known, the current biased single Josephson junction (JJ) without sufficient electrical capacitance (in other words, overdamped), does not show any chaotic behavior~\cite{kau96}. To understand this fact, consider the RCSJ model of the JJ~\cite{buc04}:
\begin{equation}
 \frac{C\hbar}{2e}\ddot{\varphi} + \frac{\hbar}{2e R_N}\dot{\varphi} + I_c \sin\varphi = I = I_{\textrm {dc}} + I_{\textrm {ac}}\sin(\Omega t),
 \label{eq1}
\end{equation}
 where $\varphi$ is the phase difference of the macroscopic wave functions for the superconductors on either side of the insulating layer, $C$ is the electrical capacitance, $R_N$ is the normal-state resistance, $I_c$ is the critical current, $I_{\textrm{dc}}$ is the direct current (dc) bias, and $I_{\textrm{ac}}$ is the amplitude of the alternating current (ac) bias, with the angular frequency $\Omega$. 
 
High damping, i.e. low $R_N$, moderates the inertial effects produced by the first term in Eq.~(\ref{eq1}). One can show that there is the threshold for the Stewart-McCumber parameter,
\begin{equation} \label{eq2}
    \beta_c = 2e I_c R^2_N C/\hbar< \frac{1}{4},
\end{equation}
below which one can disregard the inertial effects on the phase dynamics~\cite{lev88,qia88}. Formally, if (\ref{eq2}) is satisfied, then one can neglect the term $C \hbar \ddot{\varphi}/(2e)$ in (\ref{eq1}), and expect no chaotic behavior in the resulting, overdamped system, a result of the Poincar\'{e}-Bendixson theorem~\cite{str94,hil00,kal20}.

However, if additional degrees of freedom (for example, magnetic degrees of freedom) are coupled to the overdamped Josephson junction, then chaotic behavior may occur. The $\varphi_0$ junction, containing a ferromagnetic interlayer, is a good example of such coupling  (see,~\cite{shukrinov2022anomalous,Melnikov2022,bobkova2022magnetoelectric}, and the references therein).  In this type of junction, the Rashba spin-orbit coupling (SOC) within the ferromagnetic interlayer can be modeled by the Rashba Hamiltonian, $\hat{H}_R = \alpha_{R} \left[ \hat{\boldsymbol{\sigma}} \times \hat{\mathbf{p}} \right] \cdot \hat{\mathbf{n}} $, where $\alpha_{R}$ is the Rashba SOC coupling strength, $\hat{\boldsymbol{\sigma}}$ is the vector of Pauli matrices, $\mathbf{p}$ is the electron momentum and $\hat{\mathbf{n}}$ is the unit vector describing the direction of the structural anisotropy in the system. Here, we choose $\hat{\mathbf{n}} = \hat{\mathbf{e}}_z$. The electrical current through the $\varphi_0$ junction induces triplet superconducting correlations via the SOC by means of the direct magnetoelectric effect. These correlations host the electron-spin polarization which interacts with the internal magnetization of the interlayer via the exchange mechanism~\cite{bobkova2018spin,bobkova2020magnetization}. This mechanism induces motion of the magnetization via the applied current and produces new dynamical regimes which may influence the current-voltage characteristics (CVCs) of the junction \cite{Shukrinov22,shu22,Janalizadeh2022,shu23,shu20c,bot23,rabinovich2019resistive}.

The current-induced magnetization dynamics in $\varphi_0$ JJs has been studied theoretically in many previous works~\cite{shu18,shu22,shu22,bobkova2018spin,bobkova2020magnetization,shukrinov2017magnetization,shukrinov2018re,kon09,konschelle2019erratum,mazanik2020analytical,rabinovich2019resistive,Guarcello20,guarcello2021thermal,guarcello2023switching,shukrinov2019ferromagnetic,atanasova2019periodicity,Shukrinov21,Shukrinov22,Janalizadeh2022,shu23,shu20c,shu20a,shu20b,bot23,bobkov2022long,bob24,rabinovich2020electrical,nashaat2019electrical}. The variety of structures discussed in these works may be classified according to the way in which the SOC and the ferromagnetic element are combined, and through which part(s) of the structures the supercurrent flows. Such a classification leads to three essential types of $\varphi_0$ JJs: (i) superconductor/ferromagnetic metal with Rashba SOC/superconductor (S/FM with Rashba/S)~\cite{kon09,konschelle2019erratum,shukrinov2018re,mazanik2020analytical,shukrinov2017magnetization,bobkov2022long,bob24,bobkova2018spin,rabinovich2019resistive,Guarcello20,guarcello2021thermal,guarcello2023switching,shukrinov2019ferromagnetic,atanasova2019periodicity,Shukrinov21,Shukrinov22,shu22,Janalizadeh2022,shu23,shu20c,shu20b,shu20a,bot23}, (ii)  superconductor/ferromagnetic insulator/superconductor on top of the 3D topological insulator (S/FI/S on TI)~\cite{bobkova2020magnetization,rabinovich2020electrical,bobkov2022long}, and (iii) superconductor/ferromagnetic metal/superconductor on  top of the 3D topological insulator (S/FM/S on TI)~\cite{nashaat2019electrical}. For clarity, in Table~\ref{tab:data}, we provide a summary of each system type.

To our knowledge, there are no experimental studies that specifically probe the current-induced magnetization dynamics in $\varphi_0$ JJs. However, every ingredient of this problem has already been touched upon in experiments.  For example, the existence of the $\varphi_0$ phase shift has been experimentally confirmed in~\cite{assouline2019spin,zhang2022anomalous,mayer2020gate,strambini2020josephson,szombati2016josephson,murani2017ballistic}, where the anomalous phase $\varphi_0$ was achieved either due to the external magnetic in-plane field or uncontrolled magnetic impurities. Also, superconductivity on its own is well-known for its direct influence on the magnetization dynamics in S/F/S structures. In particular, the ferromagnetic resonance frequency of the magnetic interlayer appears to be greatly enhanced in S/F/S structures \cite{golovchanskiy2020magnetization,golovchanskiy2022magnetization}. This experimentally observed effect was explained by Silaev~\cite{Silaev22} in terms of the interaction between the  eddy currents induced in the superconducting leads by the interlayer dynamics and  the interlayer, itself. Here we do not take this effect into account, as it is thought to be significant only when the widths of the leads and interlayers become wide in comparison to the London penetration depth, which we assume is sufficiently large for the structures under our consideration.

In some previous works~\cite{Shukrinov22,Janalizadeh2022,shu23,shu20c,bot23,shu22}, the effect of the magnetization dynamics of monodomain interlayers on the CVCs of $\varphi_0$ JJs was investigated. In these papers, only the underdamped regime was considered for the rather high value ($\beta_c = 25$) of the Stewart-McCumber parameter. A variety of unusual features were found in the CVCs, including multiple branch structure, negative differential resistance, and chaos. The main purpose of our present work is to demonstrate that some of these features, which were previously only studied in a restricted class of underdamped dc+ac biased JJs, do indeed survive in certain overdamped dc+ac biased $\varphi_0$ JJs. Mathematically, the observation of unusual features such as chaos and multistability in overdamped dc+ac biased $\varphi_0$ JJs is only possible because the state space for the $\varphi_0$ JJs is constituted by the Josephson phase difference, $\varphi$, as well as the magnetization direction, $\mathbf{m}$, compared to only $\varphi$ for the usual overdamped dc+ac biased JJs. In the previous works~\cite{Shukrinov22,shu22,Janalizadeh2022,shu23,shu20c,bot23}, one had both the capacitance of JJs (since $\beta_c > 0$) and the magnetization dynamics, making it more difficult to show the origin of the unusual features that were found. Here, by contrast, we clearly demonstrate that it is really the current-induced magnetization dynamics which is responsible for the additional hysteretic branches and the chaotic regions in CVCs. 

The current-induced magnetization dynamics of $\varphi_0$ Josephson junctions is interesting from a fundamental perspective and for its potential applications~\cite{Guarcello20,mazanik2020analytical,bobkova2020magnetization}. The magnetization reversal procedure may be employed in such junctions to create classical bits that store information using the direction of the interlayer magnetization: one of the stable directions means $1$, while another one stands for $0$. Our present work broadens our current theoretical understanding about the magnetization dynamics in $\varphi_0$ Josephson junctions and provides an alternative way in which classical bits may be constructed. We show that switching between different branches of the current-voltage characteristics may be induced by using current pulses. In this case, the information is encoded by the voltage state of the JJ. These different voltage states exist because of the multistability in the magnetization dynamics and lead to the possibility of switching that is robust to the levels of thermal noise, estimated for real systems.

This paper is organized as follows. In Section~\ref{sec:Models}, we describe the different types of $\varphi_0$ JJ under consideration, including details of the numerical methods used in simulations and estimations of the relevant parameter ranges.  In Section~\ref{sec:Features}, we present and discuss the main results of our simulations for the different models of overdamped dc+ac biased $\varphi_0$ JJs that were formulated in Section~\ref{sec:Models}. These results include observations of multiple branches in the CVCs due to multistability. In Section~\ref{sec:Switching}, we discuss one possible application (related to switching) of the features described in Section~\ref{sec:Features}. In Section~\ref{sec:Stability}, we demonstrate that the main branch structure observed in the CVCs is robust under the influence of thermal fluctuations.  Finally, we present our conclusions in Section~\ref{sec:Conclusion}. The Appendix provides an analysis to explain the observed breaking of the one-to-one correspondence between magnetization and voltage states.

\section{Models}
\label{sec:Models}

The magnetization dynamics of the interlayer of $\varphi_0$ JJs is described by the Landau-Lifshitz-Gilbert (LLG) equation~\cite{pitaevskii2017course}:
\begin{equation} \label{eq3}
	\frac{d \bf{M}}{dt} = - \gamma \mathbf{M} \times \mathbf{H}_{\mathrm{eff}} + \frac\alpha M_0 \mathbf{M} \times \frac{d \mathbf{M}}{dt}.
\end{equation}
Here, $\gamma$ is the gyromagnetic ratio, $\mathbf{M}$ is the interlayer magnetization, $M_0 = \vert \mathbf{M} \vert$, and $\alpha$ is the Gilbert damping constant. We do not distinguish between the magnetic anisotropy field of the interlayer and the torque exerted on the magnetization. Both these quantities are included in the term $\mathbf{H}_{\mathrm{eff}}$. For simplicity, we will consider only the easy-axis anisotropy along $\hat{\mathbf{e}}_z$. For such an interlayer, the magnetic energy is $E_M = -K \mathcal{V} M^2_z/(2  M^2_0)$, where $K$ is the magnetic anisotropy constant and $\mathcal{V}$ is the volume of the ferromagnetic interlayer.

\begin{table*}[t!]
\caption{\label{tab:data} The three system/model types under consideration.}
\begin{tabular}{|m{0.28\linewidth}|m{0.29\linewidth}|m{0.4\linewidth}|}
\hline \vspace*{1mm}
\textbf{ Type (i): S/FM with SOC/S} &  \phantom{AAA}\textbf{Type (ii): S/FI/S on TI} & \phantom{AAAAA} \textbf{Type (iii):  S/FM/S on TI } \\[1mm]
    \hhline{|=|=|=|}
   \begin{minipage}{0.99\linewidth}
        \centering
        \includegraphics[height=2.5cm]{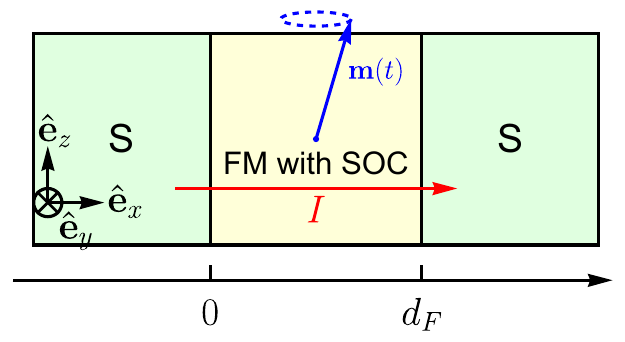}
   \end{minipage}   & \begin{minipage}{0.99\linewidth}
        \centering
        \includegraphics[height=3.4cm]{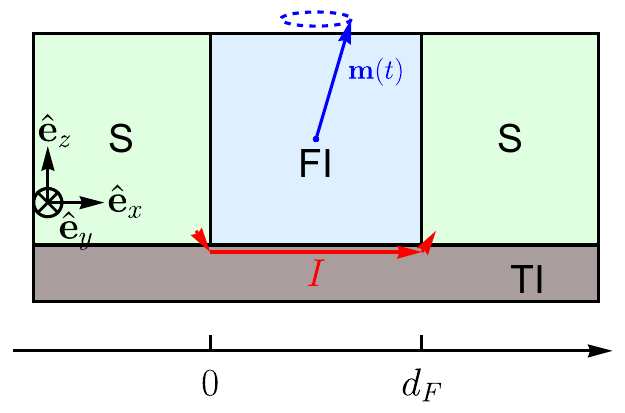}
   \end{minipage}  & \begin{minipage}{0.99\linewidth}
        \centering
        \includegraphics[height=3.4cm]{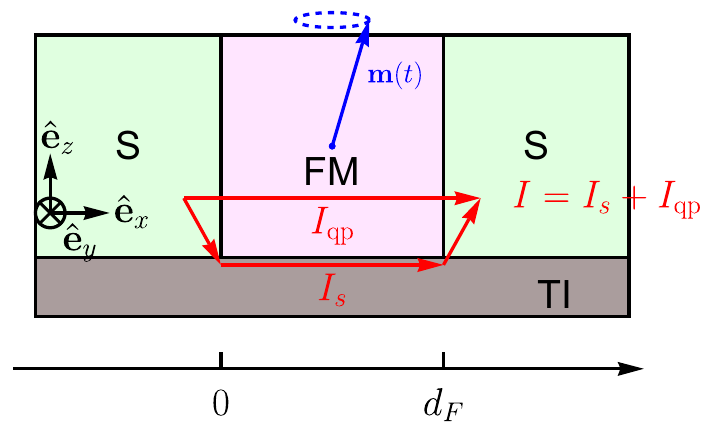}
   \end{minipage}   \\
   \hline
   $\begin{aligned}  & \\
            & \phantom{AA} I_c(m_x) = I_{c0}  = \textrm{const}, \\[6mm]
            & \phantom{AA} I_{s} = I_{c0} \sin(\varphi - r m_y).
            \\
            & \phantom{A}
    \end{aligned}$
   & \multicolumn{2}{c|}{ $\begin{aligned} & \\
            &I_c(m_x) = I_{c0}\mathcal{N}^{-1}\int_{-\pi/2}^{\pi/2}\mathrm{d}\phi \cos\phi \exp{\left(-\frac{2\pi k_B T_c d_F}{\hbar v_F \cos\phi}\right)} \cos\left(r m_x\tan\phi\right) = I_{c0}i_{c}(m_x),\\[3mm]
            &I_{s} = I_c(m_x)\sin(\varphi - r m_y),\ \mathcal{N} = \int_{-\pi/2}^{\pi/2}\mathrm{d}\phi \cos\phi \exp{\left(-\frac{2\pi k_B T_c d_F}{\hbar v_F \cos\phi}\right)}, \ f(m_x) = \frac{\partial i_{c}(m_x)}{\partial m_x} \\
            & \phantom{A}
    \end{aligned}$}\\
   \hline
   $\begin{aligned}
    & \mathbf{H}_{\mathrm{eff}1} = \mathbf{T}_1 + m_z \hat{\mathbf{e}}_z,\\
    \phantom{AA} \mathbf{T}_1  & = Gr\frac{I_{s}}{I_{c0}} \hat{\mathbf{e}}_y \\
    & = Gr \sin\left(\varphi - rm_y\right)\hat{\mathbf{e}}_y . \\
    & \phantom{A}
    \end{aligned}
   $& $\begin{aligned}
    & \mathbf{H}_{\mathrm{eff}2} = \mathbf{T}_2 + m_z \hat{\mathbf{e}}_z,\\
   \phantom{AA} \mathbf{T}_2 & = Gr\frac{I_{s} + I_{\mathrm{qp}}}{I_{c0}}\hat{\mathbf{e}}_y \\  
   & = Gr \left(i_{\textrm{dc}} + i_{\textrm{ac}}\sin\left(\Omega t\right)\right)\hat{\mathbf{e}}_y. \\
   & \phantom{A}
   \end{aligned}$  &
   $\begin{aligned} 
     & \\
     & \mathbf{H}_{\mathrm{eff}3} =  \mathbf{T}_3 + m_z \hat{\mathbf{e}}_z, \\
     \mathbf{T}_3  & = Gr\left\{f(m_x)\left(1-\cos\left(\varphi - rm_y\right)\right)\hat{\mathbf{e}}_x + \frac{I_{s}}{I_{c0}}\hat{\mathbf{e}}_y\right\} \\
    & =  Gr\left\{ f(m_x) \left(1 - \cos\left(\varphi - r m_y\right)\right) \hat{\mathbf{e}}_x  \right. \\ & \qquad \! \; \; \; \; \;  \left. + \,  i_c(m_x) \sin\left(\varphi - r m_y\right) \hat{\mathbf{e}}_y\right\}. \\
    & \phantom{A}
   \end{aligned} $ \\
   \hline
\end{tabular} 
\end{table*}

As mentioned in the Introduction, there are essentially three types of $\varphi_0$ JJs, differing only by the  applied effective magnetic fields in the interlayer, and the current-phase relationships. Physically, the differences between effective magnetic fields are related to how the total current flowing through the system gets divided into the superconducting current $I_{s}$ and the quasiparticle current $I_{\textrm{qp}}$. According to the modified resistivily shunted junction (RSJ) model without capacitance~\cite{rabinovich2019resistive,rabinovich2020electrical}, we may write:
\begin{equation} \label{eq4}
\begin{aligned}
	&I = I_{\textrm{dc}} + I_{\textrm{ac}}\sin \left( \Omega t \right)  = I_{s} + I_{\textrm{qp}},\\
	&I_{s} =  I_c (m_x)\sin{(\varphi - \varphi_0)}\text{, }
	I_{\textrm{qp}} =  \frac{\hbar}{2 e R_N}\left( \dot{\varphi} - \dot{\varphi}_0\right).
\end{aligned}
\end{equation}
Here $\varphi_0$ is the anomalous phase shift in the junctions, $\mathbf{m}= \mathbf{M}/M_0$. In all these types of JJs we have $\varphi_0 = r m_y$, where the dimensionless constant $r \sim \alpha_R$ characterizes the intensity of the SOC.  The effective magnetic fields $\mathbf{H}_{\mathrm{eff}1,2,3}$, which are used for the calculation of the magnetization motion, contain torques $\mathbf{T}_{1,2,3}$, which describe the effect of the current on the magnetization. 

The type (i) model takes into account only the superconducting part of the current, $I_{s}$, in the torque acting on the magnetization, $\mathbf{T}_{1} \propto I_{s} \hat{\mathbf{e}}_y$. The quasiparticle current, in principle, may induce the electron-spin polarization and, therefore, participate in the  formation of the torque $\mathbf{T}_1$ \cite{bobkov2022long,bob24,rabinovich2019resistive},  but we neglect it for the consistency with the previous works \cite{kon09,konschelle2019erratum,shukrinov2018re,mazanik2020analytical,shukrinov2017magnetization,bobkova2018spin,Guarcello20,guarcello2021thermal,guarcello2023switching,shukrinov2019ferromagnetic,atanasova2019periodicity,Shukrinov21,Shukrinov22,Janalizadeh2022,shu23,shu20c,shu20b,shu20a,bot23,shu22,shu18}. If we included the quasiparticle contribution in the torque $\mathbf{T}_1$, then we would have an additional term $ \sim r I_{\textrm{qp}}$ in $\mathbf{T}_1$. This term only influences the magnetization dynamics and does not affect the JJ phase evolution, expressed through $\Phi = \varphi - r m_y$. As we shall see in Sec.~\ref{sec:Features}, the CVCs of the type (i) junctions are independent of the magnetization, since their critical currents are independent of the magnetization direction. Therefore, the inclusion of the quasiparticle contribution in $\mathbf{T}_1$ would not change any of the results for the type (i) model.

The type (ii) model uses the total current flowing through the Josephson weak link which is organized through the surface states of the TI, so $\mathbf{T}_{2} \propto I = \left(I_{\textrm{dc}} + I_{\textrm{ac}}\sin(\Omega t)\right)\hat{\mathbf{e}}_y$ \cite{bobkova2020magnetization,bobkov2022long,rabinovich2020electrical} due to the spin-momentum locking in the TI. 

The type (iii) model assumes that the quasiparticle current flows through the FM layer while the superconducting current flows through the Josephson weak link formed by the surface states of the TI, i.e., only the superconducting part of the total current is used in the torque: $\mathbf{T}_{3} \sim I_{c0} f(m_x)\left(1 - \cos\left(\varphi - r m_y\right)\right) \hat{\mathbf{e}}_x + I_{s} \hat{\mathbf{e}}_y$, due to the spin-momentum locking in the TI~\cite{nashaat2019electrical}. The formulas for $i_c(m_x)$ and $f(m_x)$ are written in Table~\ref{tab:data}. The difference between (ii) and (iii) types of $\varphi_0$ JJs occurs because the resistance of the FM layer is typically much smaller ($\sim 0.1 \; \mathrm{to} \; 1\ \Omega$, \cite{khaire2009critical}) than the resistance of surface states of the TI ($\sim 50\ \Omega$, \cite{oostinga2013josephson,veldhorst2012josephson}), and so, the quasiparticles 'prefer' to flow through the ferromagnetic metal. Models (ii) and (iii) are thus distinguished by the different currents flowing through that part of each structure hosting the SOC and electron-spin polarization.

\begin{figure}[htb!]
    \centering
    \includegraphics[width=0.35\textwidth]{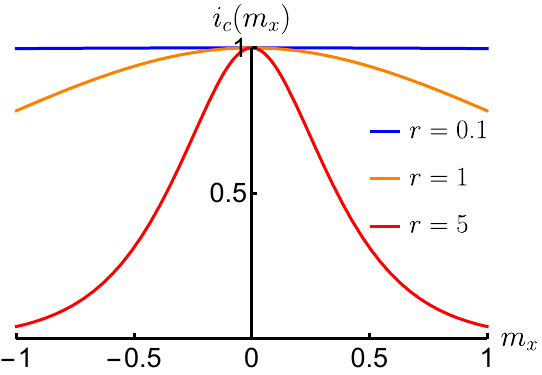}
    \caption{The dependence of the critical current on the magnetization direction $i_c(m_x) = I_c(m_x)/I_{c0}$ from Table~\ref{tab:data} for different $r$. Here we use $I_{c0} = I_c(m_x = 0)$. The dimensionless length is $\tilde{d}_F = 2\pi k_B T_c d_F/(\hbar v_{F}) = 1$.}
    \label{fig1}
\end{figure}

Model (i) differs crucially from (ii) and (iii) due to the different dependence of the critical current on the magnetization direction $I_c(m_x)$. As stated in Table~\ref{tab:data}, in model (i) the critical current is independent of $\mathbf{m}$, i.e. $I_c(m_x) = I_{c0}= \mathrm{const}$, while in models (ii) and (iii) it depends on $m_x$ as shown in Fig.~\ref{fig1}.  
The function $I_c(m_x)$ depends on the SOC parameter $r$ and the length of the junction $d_F$ in the current direction $\hat{\mathbf{e}}_x$, given in dimensionless form by ${\tilde{d}_F = 2\pi k_B T_c d_F/(\hbar v_F)}$, where $k_B$ is the Boltzmann constant, $T_c$ is the critical temperature of the superconductors and $v_F$ is the Fermi velocity of the TI surface states. In principle, model (i) may also have a non-constant $I_c(m_x)$ dependence~\cite{Hasan2022}, however, this dependence would be very slight due to the small Rashba SOC in ferromagnetic metals~\cite{Greening2020,mir10}: $r \sim 0.1$ for $\alpha_R \sim 10^{-3} \; \mathrm{to} \;  10^{-1} \lesssim 1 \; \mathrm{eV} \, \angstrom$. Rashba SOC intensities $\alpha_R \sim 1\ \mathrm{eV} \, \angstrom$ have been achieved in Co/Pt thin films with Co layers $\sim 0.6\; \mathrm{nm}$~\cite{mir10}. For such films, $r$ is also small ($r \sim 0.01)$, making $I_c(m_x) \approx \mathrm{const}$ a good approximation. In comparison, for TIs~\cite{assouline2019spin}, $r \sim 1 \; \mathrm{to} \; 10$ and $\alpha_R \sim 1 \; \mathrm{to} \; 3 \; \mathrm{eV} \ \angstrom$.

To calculate numerically the CVCs, the magnetization dynamics and the Lyapunov exponents (LEs) from the system of Eqs.~(\ref{eq3}) and (\ref{eq4}), we first re-write the equations in terms of the dimensionless variables:
\begin{equation} 
\begin{aligned}
    & t \to \omega_F t,\ \omega_F = \gamma K/M_0,\ \Omega\to \Omega/\omega_F;\\
    &\omega_c = 2 e I_{c0}R_{N}/\hbar \to \omega_c/\omega_F;\\
    & I_c(m_x) = I_{c0} i_c(m_x)\text{ with } I_{c0} = I_c(m_x = 0);\\
    & i = \frac{I}{I_{c0}} ,\ i_{\textrm{dc}} = \frac{I_{\textrm{dc}}}{I_{c0}},\ i_{\textrm{ac}} = \frac{I_{\textrm{ac}}}{I_{c0}};\\
    & G =  \frac{E_J}{K\mathcal{V}} = \frac{ \hbar I_{c0}}{ 2 e K d_F d_z d }, \;  w = \frac{\omega_F}{\omega_c}.
\end{aligned}
\label{eq5}
\end{equation}
As before, $d_F$ is the length of the junction along the current direction $\hat{\mathbf{e}}_x$ and $d$ is its width in the $\hat{\mathbf{e}}_y$ direction. Here we assume that the junction length along the $\hat{\mathbf{e}}_z$ direction is $d_z$, so that the volume of the ferromagnetic interlayer can be written as $\mathcal{V} = d_F d_z d$. With the help of Eqs.~(\ref{eq5}), we can then write Eqs.~(\ref{eq3}) and (\ref{eq4}) in dimensionless form as:
\begin{subequations} 
\label{eq6}
\begin{eqnarray}
\dot{\mathbf{m}} & = &  - \frac{1}{1+\alpha^2}\left\{\mathbf{m}\times
\left( \mathbf{T}_{i}+m_{z}\mathbf{\hat{e}}_{z}\right) + \right.   \nonumber \\ 
& & \hspace*{1.2cm} \left.   +\alpha \mathbf{m\times }\left[ \mathbf{m}\times \left( \mathbf{T
}_{i}+m_{z}\mathbf{\hat{e}}_{z}\right) \right]\right\},  \label{eq6a}\\
w\dot{\Phi} & = &   i_{\textrm{dc}} + i_{\textrm{ac}}\sin \left( \Omega
t\right) - i_{c}\left( m_{x}\right) \sin \Phi .  \;   \label{eq6b}
\end{eqnarray}
\end{subequations}
The torques $\mathbf{T}_{i} \; (i = 1,2,3)$, appearing in Eq.~(\ref{eq6a}), are given in dimensionless form in Table~\ref{tab:data}. For convenience, we have introduced the new variable, $\Phi = \varphi - rm_y$, in Eq.~(\ref{eq6b}), which has the same form as that of an ordinary, overdamped, ac-driven JJ. The average dimensionless voltage for any given dc current $i_{\textrm{dc}}$ and any fixed ac current $i_{\textrm{ac}}$ is calculated as $V/(I_{c0}R_N) = w \overline{\dot{\varphi}} = w \frac{\varphi(T) - \varphi(0)}{T}$. We denote the time average of any time-dependent quantity such as $\dot{\varphi}(t)$, over the time domain $t \in [0,T]$, by $\overline{\dot{\varphi}}$. Typically, in the numerical results that follow we allow a transient time of $500$ drive cycles (of period $\tau = 2\pi/\Omega$), before averaging over $T=500\tau$ for the CVCs, and $T = 128 \, 000$, for the LEs. The LEs are calculated via the full spectrum method~\cite{shi79,ben80,wol85}, using the explicit pseudosymplectic method~\cite{daq18} (for further details, see also the Appendix of~~\cite{bot23}).

Experimentally, the system parameters fall into the following ranges. The Gilbert damping $\alpha$ is typically much smaller than one~\cite{mazanik2020analytical,bobkova2020magnetization}: ${\alpha \sim 10^{-4} \; \mathrm{to} \; 10^{-2}}$. The coefficient $G$ (see Eq.~\ref{eq5}) has been estimated in the range~\cite{kon09,nashaat2019electrical,bobkova2020magnetization}: ${ G \sim 10^{-2} \; \mathrm{to} \; 10^{2}}$. The following numerical parameters were used for this estimate: ${d \sim 100 \; \textrm{nm}}$, ${d_F \sim 10 \; \textrm{nm}}$, ${d_z \sim 5 \; \textrm{nm}}$, $I_{c0} \sim 6~\mathrm{\mu A}$, ${K \sim 1}$ to $10^{4}~\textrm{J}\, \textrm{m}^{-3}$. The critical current density of $60~\mathrm{A/m}$, used here, is consistent with~\cite{veldhorst2012josephson,oostinga2013josephson}, where Nb/Bi$_2$Te$_3$/Nb and Nb/HgTe/Nb junctions were investigated. The range of the magnetic anisotropy coefficient, $K \sim 10^1$ to $10^{4} \; \mathrm{J} \, \mathrm{m}^{-3}$ can be achieved in different magnetic bilayers~\cite{rus04,bal21,mendil2019magnetic}. The SOC coefficient $r$ has been estimated in~\cite{kon09,rabinovich2020electrical,bobkova2020magnetization} to be in the range ${r \sim 0.1 \; \mathrm{to} \; 10}$. The small value of the SOC intensity $r \ll 1$ is typical for JJs with FM with the SOC interlayer like permalloy doped with Pt atoms \cite{shukrinov2018re,Greening2020}. The smallness of the SOC intensity $r \ll 1$ justifies the approximation of the critical current independence on the magnetization direction, see Fig.~\ref{fig1}. The high value of the SOC intensity $r\sim 10$ may exist in JJs based on the TI surface states \cite{bobkova2020magnetization,rabinovich2020electrical,bobkov2022long}.  So, we expect that both $G$ and $r$ may vary over wide ranges, depending on the quality of the experimentally made structures. With the help of these values, $\omega_F = \gamma K/M_{0}  \approx 10^{10} \; \mathrm{Hz}$ \cite{kon09,golovchanskiy2020magnetization,golovchanskiy2022magnetization}. Typically, for the type (ii) systems we have $R_N \sim 50\ \Omega$ \cite{oostinga2013josephson,veldhorst2012josephson}, stemming from the resistance of the surface states of TI, and the parameter $w$ may be estimated as $w \sim 10^{-2}$, which is consistent with the value used in~\cite{mazanik2020analytical}. However, the parameter $w$ can be different for the model (iii) because in that case $R_N$ is determined by the resistance of the ferromagnetic metal which may be rather small $R_N \sim 1\ \Omega$ \cite{khaire2009critical}, so in that case $w \sim 1$.

It is difficult to estimate the typical capacitance of systems under consideration, but there are experimental works which show that it is possible to achieve non-hysteretic CVCs in JJs based on ferromagnetic interlayers or TI \cite{khaire2009critical,oostinga2013josephson}. For example, in the experimental work~\cite{oostinga2013josephson}, the Stewart-McCumber parameter was estimated to be $\beta_c \sim 0.001$ for the JJ based on the TI, HgTe. Generally, the overdamped approximation, as in Eqs.~(\ref{eq2}) and (\ref{eq4}), applies to systems in which there is no too much self-heating~\cite{courtois2008origin}.

\section{Features in the CVCs} \label{sec:Features}
We now proceed to the main results. It is well-known that a magnetic nanoparticle, which experiences a constant longitudinal and a time-dependent transverse components of the external magnetic field or a time-dependent voltage-controlled magnetic anisotropy, can exhibit chaotic motion and multistability~\cite{bragard2011chaotic,velez2020periodicity,contreras2022voltage,wang2016nonlinear,wang2018emergence,nashaat2022bifurcation}. Here, for the type (i)--(iii) systems, we have a similar situation in which the magnetic interlayers of the junctions are influenced by the time-dependent ac drive. What are the consequences of this interaction for the CVCs of such JJs? We will show that, due to the magnetization dynamics, chaotic and hysteretic behaviour may also occur.

\subsection{Type (i) systems}
To begin with, we have established that the type (i) systems do not show any new dynamics in their CVCs, compared to the superconductor/insulator/superconductor (S/I/S) junctions.

To understand why the CVCs are the same, we notice that the solutions, $\Phi = \varphi - r m_y$, of the RSJ Eqn.~(\ref{eq6b}), with $I_c(m_x) = I_{c0} = \textrm{const}$, do not depend on the magnetization dynamics, but only on $i_{\textrm{dc}}$ and $i_{\textrm{ac}}$. This means that, for any $\mathbf{m}(t)$ at fixed $i_{\textrm{dc}}$ and $i_{\textrm{ac}}$, we obtain the same solutions, $\Phi$, as for the corresponding  S/I/S junctions. This leads to the voltage $V/(I_{c0} R_N) = w \overline{\dot{\varphi}} = w \overline{\left(\dot{\Phi}  + r \dot{m}_y\right)} = w \overline{\dot{\Phi}}$, because ${\overline{\dot{m}}_y = \frac{1}{T}\int_{0}^{T}\mathrm{d}t\ \dot{m}_y = \frac{m_y(T) - m_y(0)}{T} \xrightarrow[\infty]{T} 0}$.  So, the CVCs, for $\varphi_0$ JJs with $I_c(m_x) = I_{c0} = \textrm{const}$ coincide with the CVCs of the corresponding S/I/S junctions. Furthermore, the effect of the magnetization dynamics on  such $\varphi_0$  S/F/S junctions have been discussed recently~\cite{Shukrinov22,shu22,Janalizadeh2022,shu23,shu20c,bot23}. We stress here that the  dynamical aspects discussed in these previous works relied on having capacitance in the system, while in our present work, there is no capacitance.

\subsection{Type (ii) systems}
In the type (ii) systems the magnetic dynamics is decoupled from phase dynamics, but not the other way around~\cite{bobkova2020magnetization,bobkov2022long,rabinovich2020electrical}. In this case the magnetization `feels' only the total current ${i = i_{\textrm{dc}} + i_{\textrm{ac}} \sin\left( \Omega t \right)}$ and  is unaffected by the value of $\varphi$ or $\dot{\varphi}$. If we measure the CVCs of such junctions for $i_{\textrm{ac}} = 0$, we do not see any interesting effects, because the magnetization lines up along one of the stable directions: $\mathbf{m} = (0,\ Gri_{\textrm{dc}},\ \pm\sqrt{1 - (Gri_{\textrm{dc}})^2})$, for $Gri_{\textrm{dc}} < 1$, or $\mathbf{m} = (0,\ \pm1,\ 0)$ for $Gri_{\textrm{dc}} \ge 1$. These stable directions preserve $m_x = 0$, which leads to the preservation of $i_c(m_x)$ along the CVCs. Hence, since the CVCs are determined by $\varphi$ and $i_c(m_x(t))$, they are unaffected by the magnetization motion in this case.

\begin{figure}[hb]
    \centering
    \includegraphics[width=0.48\textwidth]{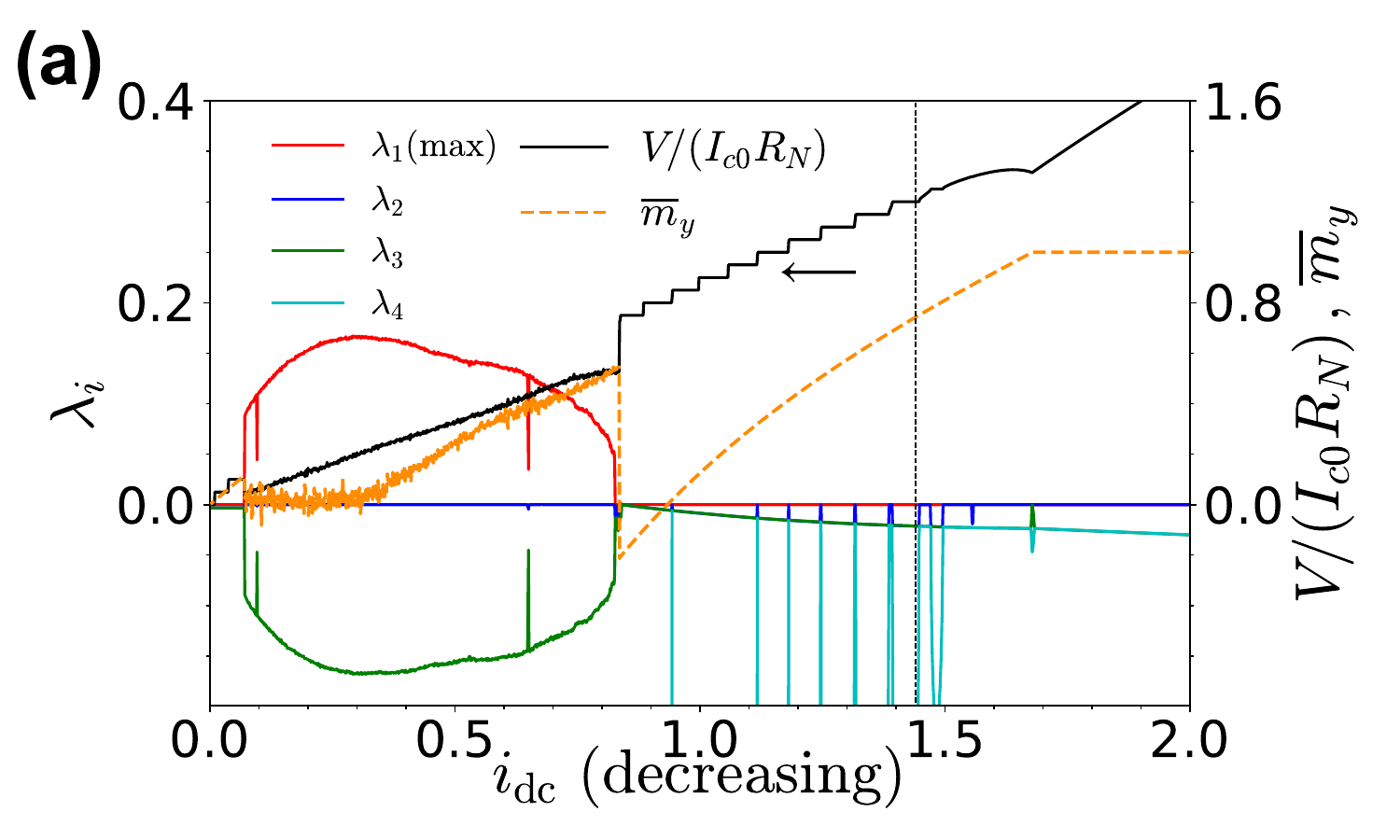}
    \includegraphics[width=0.48\textwidth]{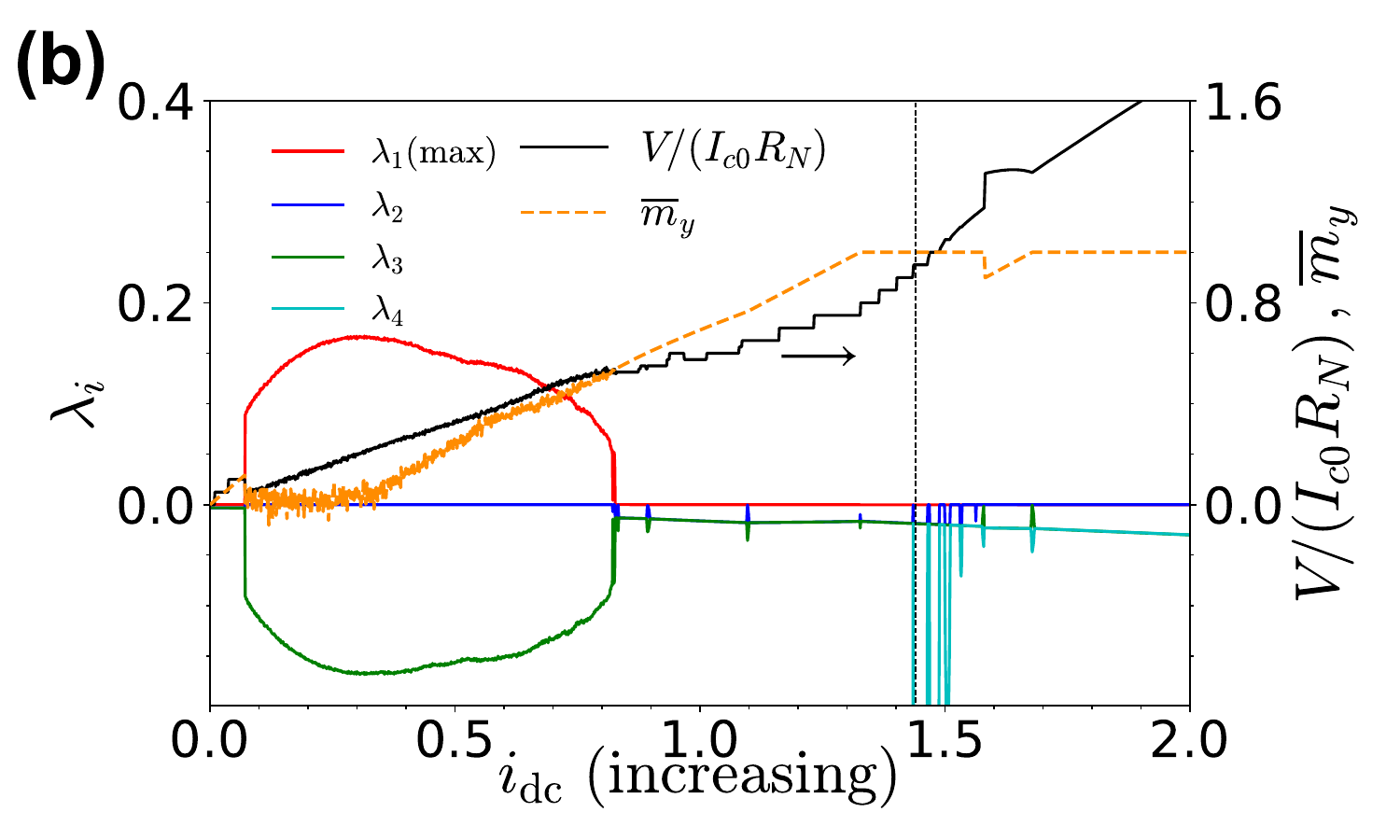}
    \caption{The typical CVCs for the model (ii), together with the LE spectrum ($\lambda_i$, $i = 1,2,3,4$), time averaged voltage $V/(I_{c0}R_N)$ and time averaged $\overline{m}_y$, shown as functions of the dc-bias $i_{\textrm{dc}}$, for (a) decreasing and (b) increasing current. The parameters of the system are $G = 0.2$, $r = 5$, $\alpha = 0.02$, $\Omega = 1$, $i_{\textrm{ac}} = 0.5$, $w = 0.05$, and $\tilde{d}_F =1$. The arrows indicate the direction of the current sweep, while the vertical grid lines correspond to $i_{\textrm{dc}} = 1.44$, a value that will be discussed in connection with Figs.~\ref{fig3} and \ref{fig8}-\ref{fig11}.}
    \label{fig2}
\end{figure}

When we consider $i_{\textrm{ac}} > 0$, for the type (ii) systems, we do find interesting features in the CVCs. In Figs.~\ref{fig2} (a) and (b), for example, we show the CVCs and the Lyapunov exponents (LE) spectrum for downward and upward sweeps of the dc-bias, respectively. With the ac drive on, the effective dimension of the autonomous state space is now four, i.e., two dimensions for the magnetic subsystem, one for the time dependent drive and one for the overdamped JJ. The LE spectrum thus consists of four exponents, one being trivially zero, since it corresponds to the perturbation tangent to the trajectory, which fluctuates with no net growth or decay~\cite{pik16}. We see the appearance of a chaotic region for $0.065 < i_{\textrm{dc}} < 0.83$, as indicated by the positive maximal LE, $\lambda_1 > 0$. In addition, we see that the $V(i_{\textrm{dc}})/(I_{c0}R_N)$ functions take different values depending on whether the dc current is increasing or decreasing. 

The appearance of the chaotic regions in the CVCs, shown in Figs.~\ref{fig2} (a) and (b), is directly related to the chaotic magnetization dynamics. To demonstrate this relationship we have also calculated the LEs for the three-dimensional magnetic subsystem on its own, i.e. the driven LLG equation~\footnote{Chaotic regimes for the magnetization dynamics of the driven LLG equation on its own are of course well-known in the literature. See, for example,~\cite{bragard2011chaotic,nis15,nis18,wang2016nonlinear,wang2018emergence,velez2020periodicity,contreras2022voltage,nashaat2022bifurcation}.}. We do not show the result of this calculation here, because the three exponents of the magnetic subsystem alone turn out to be exactly the same as the three largest exponents shown in Figs.~\ref{fig2}, i.e., the same as for the unidirectionally coupled system. From this coincidence of the LEs, we conclude that the chaos in the magnetic subsystem is a result of the ac drive, while the chaos in the voltage states of the junction merely reflects the chaotic motion inherent in the magnetic subsystem.

\begin{figure}[htbp]
    \includegraphics[width=0.48\textwidth]{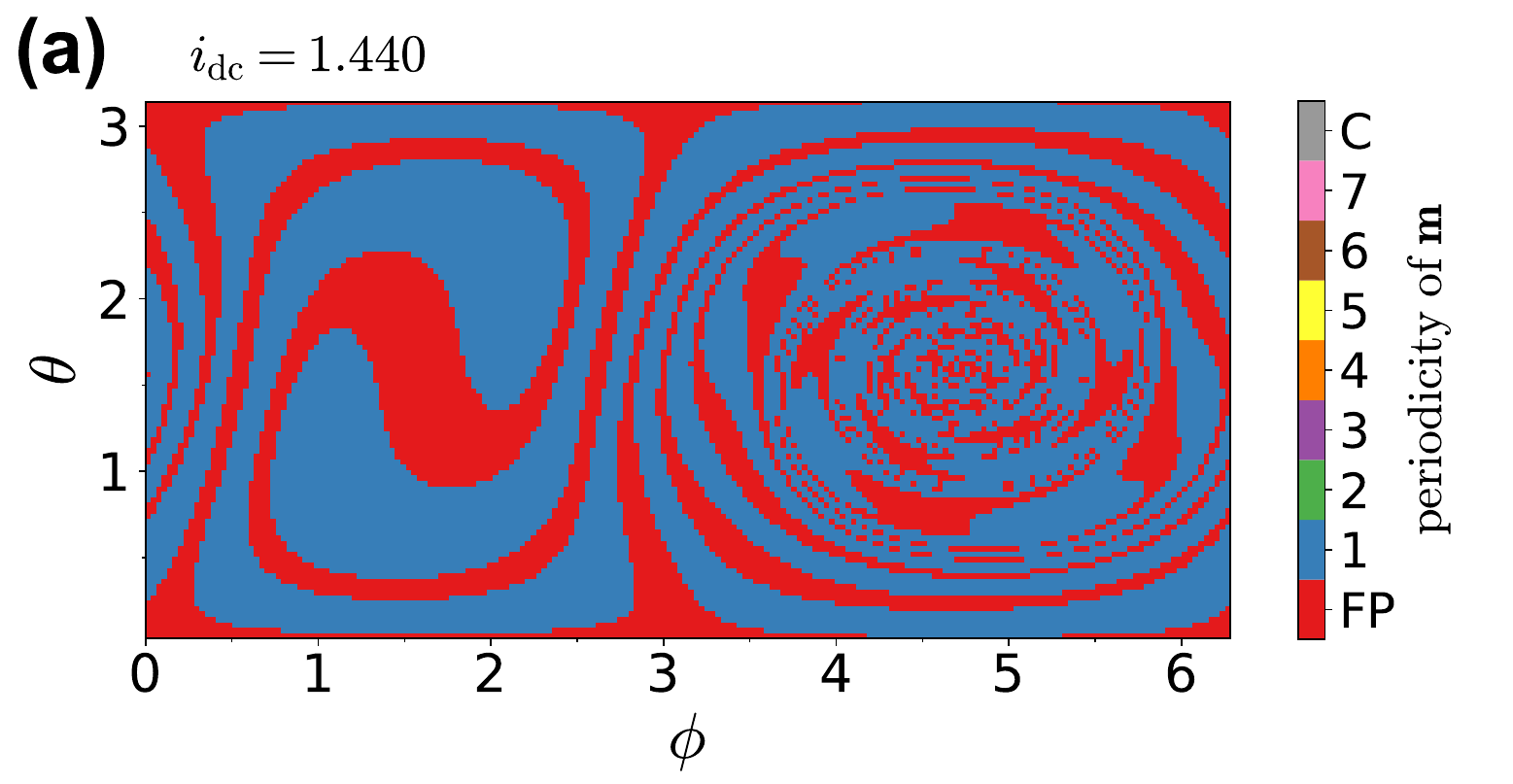}
    \includegraphics[width=0.48\textwidth]{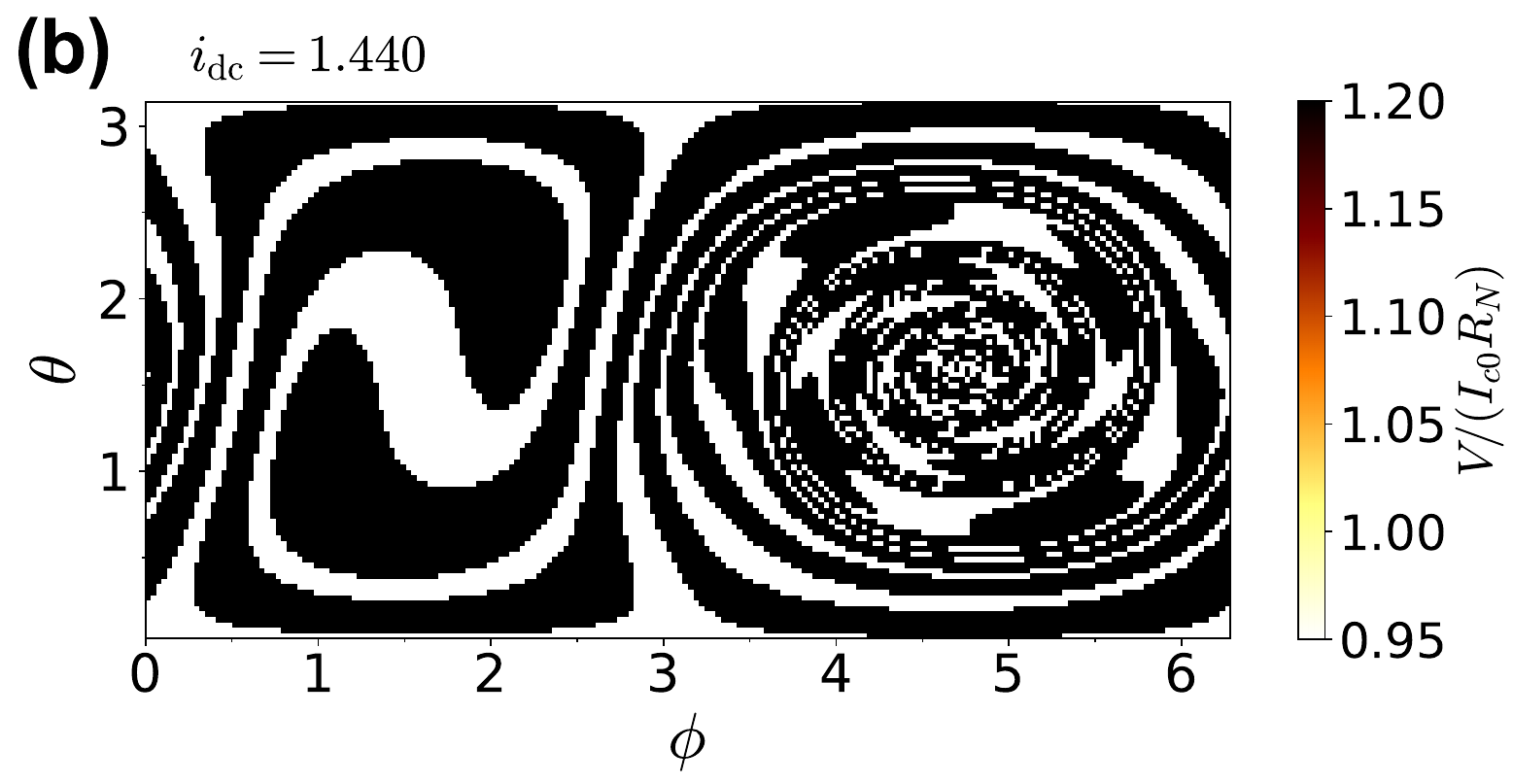}
    \caption{(a) Comparison between the basin of attraction to different periodicities in the magnetization dynamics. The basin of attraction consists of either period 1 or fixed point (FP) behaviour, with no chaos (C) in this case. See, the color scale on the right. (b) The basin of attraction to the time averaged voltage. Comparison of (a) and (b) shows that there is one-to-one correspondence between the two basins, as discussed in the main text. In both figures $\varphi=0$ at $t=0$ and $i_{\rm dc}=1.440$. All other parameters are the same as in Fig.~\ref{fig2}. } \label{fig3}
\end{figure}

The two branches in the CVCs shown in Figs.~\ref{fig2} (a) and (b) are not due to inertial effects, as there are no inertial terms in our models. Rather, as we have mentioned, they are due to different modes that occur in the magnetic subsystem. One can see the effect of the magnetization dynamics by making a comparison between the basins of attraction to different periodicities in $\mathbf{m}(t)$, relative to the drive cycle, and those for the average voltage, as shown in Fig.~\ref{fig3}. 
To obtain Fig.~\ref{fig3}, we set $\varphi=0$ at $t=0$, and solve Eqs.~(\ref{eq6}) for $19 \ 802$ initial conditions (ICs) for $m_x$, $m_y$ and $m_z$. The ICs for the magnetization are expressed in terms of the spherical polar angles $(\theta,\phi)$, and are uniformly spread over the unit sphere. Each point on the sphere is spaced at an angle of $\pi/100$, giving $99 \times 200 + 2 = 19 \ 802$ points, in total. The two extra points are added for the north $(0,0,1)$ and south $(0,0,-1)$ poles. From these initial angles, we then generate the corresponding initial Cartesian components via the usual transformations: $m_x = \sin\theta\cos\phi$, $m_y = \sin\theta\sin\phi$, and $m_z = \cos\theta$. 
\begin{figure}[htbp]
    \includegraphics[width=0.48\textwidth]{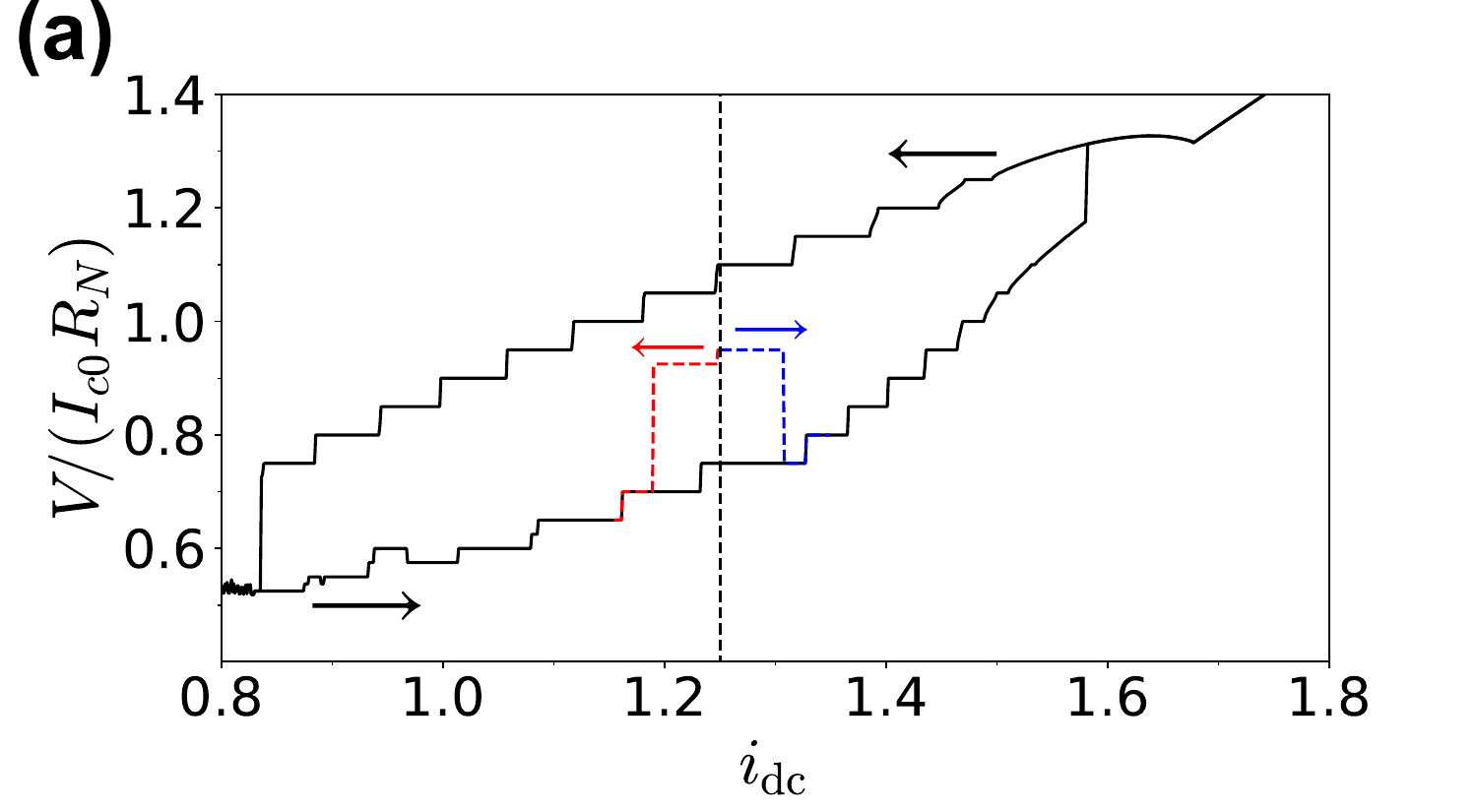}
    \includegraphics[width=0.48\textwidth]{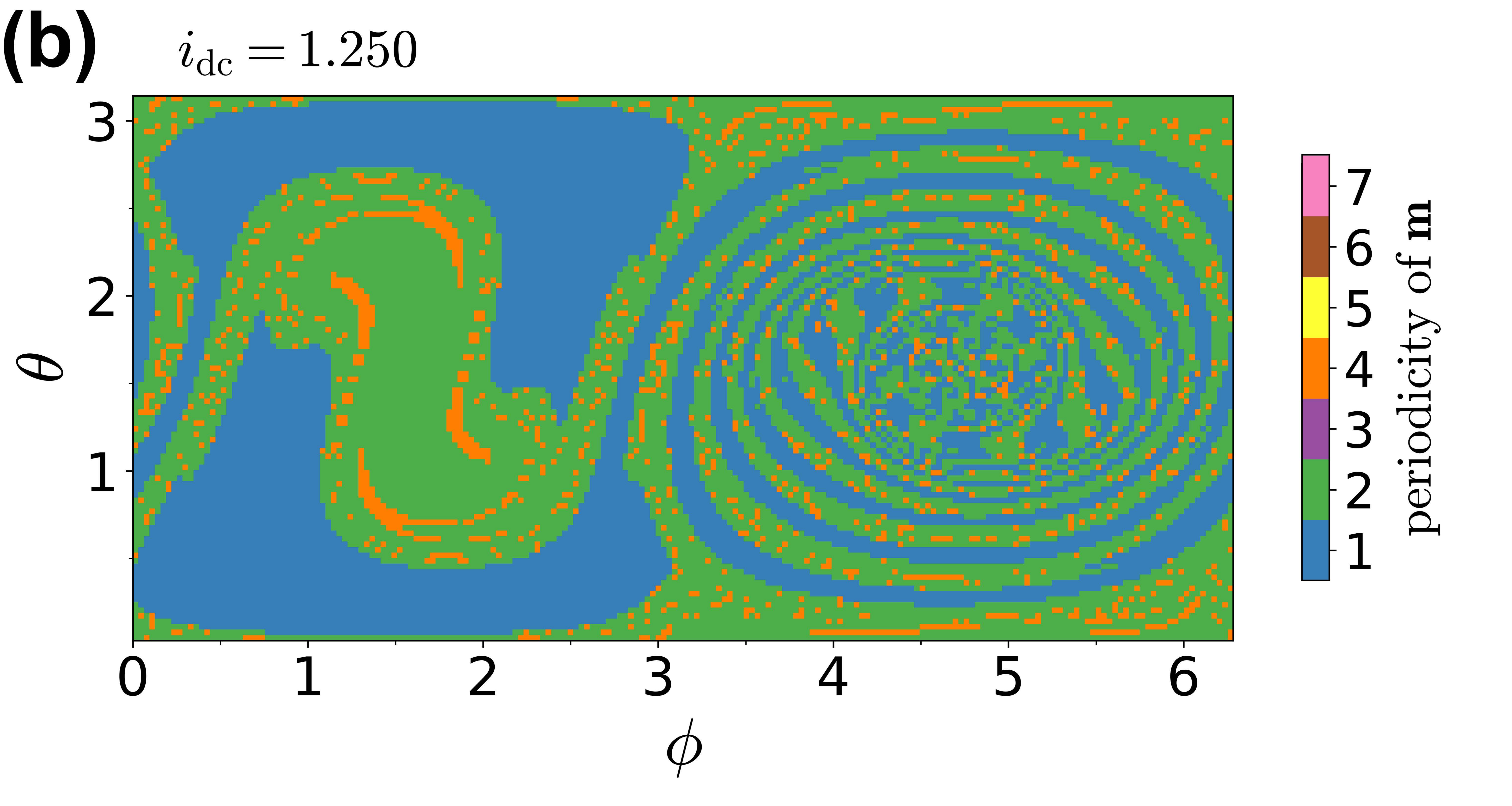}
    \includegraphics[width=0.48\textwidth]{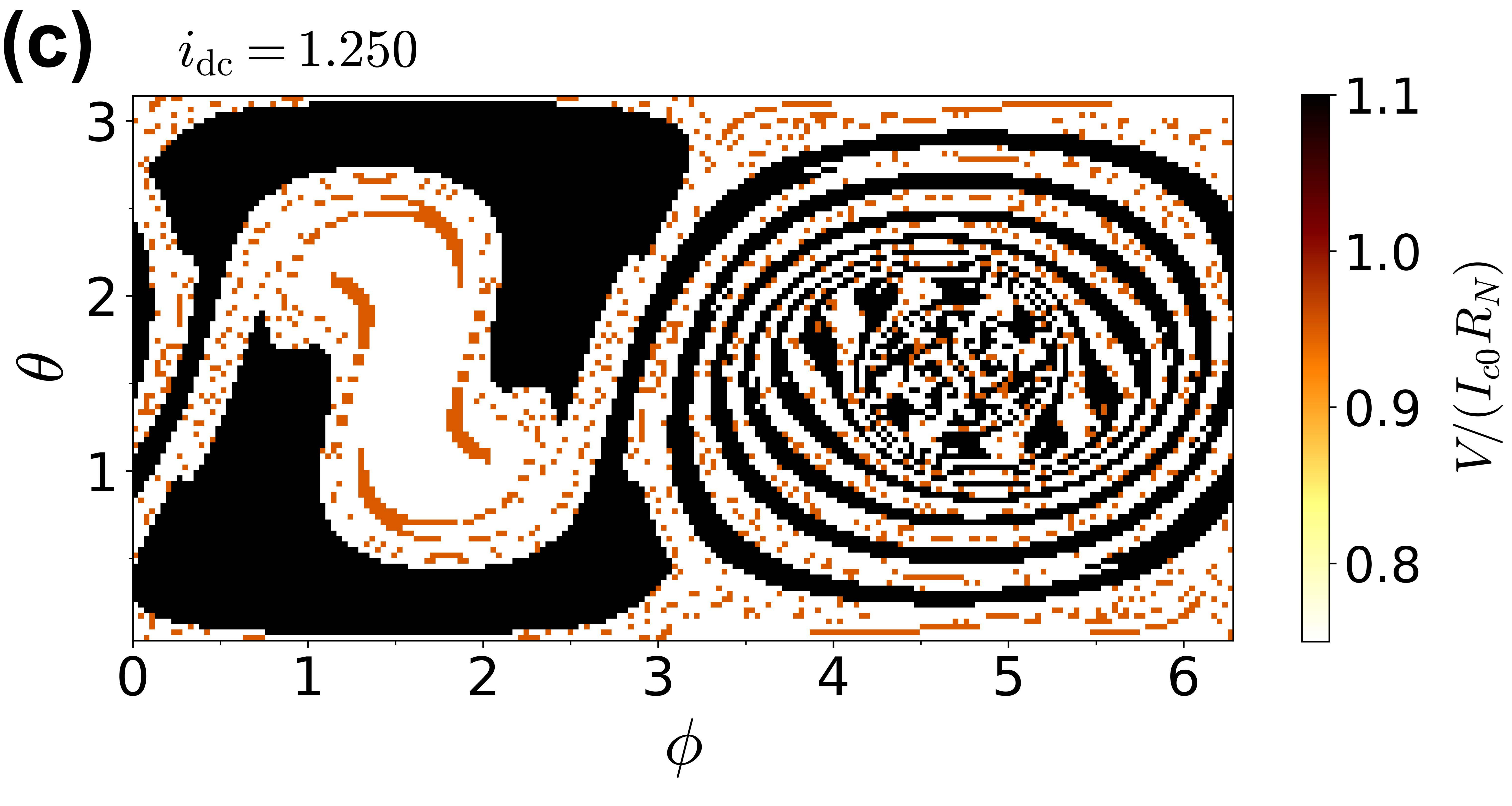}
    \caption{(a) Additional branches in the CVCs due to different magnetization modes in the type (ii) system. The parameters are the same as in Figs.~\ref{fig2} to \ref{fig3}, with the arrows indicating the sweep directions of the dc-bias $i_{\textrm{dc}}$. The black branches are the same as in Figs.~\ref{fig2} (a) and (b). The additional branches, shown by the dashed lines, were obtained by choosing a particular IC, at $i_{\textrm{dc}} = 1.25$, and then sweeping $i_{\textrm{dc}}$ up (blue dashed line) and down (red dashed line). In both directions we find period 4 behavior in $\textbf{m}(t)$. The vertical dashed grid line just shows the position of $i_{\textrm{dc}} = 1.25$. In (b) and (c), the basins of attraction corresponding to $i_{\textrm{dc}} = 1.25$ are shown for the periodicity of $\textbf{m}(t)$ and the time averaged voltage, respectively.}
    \label{fig4}
\end{figure}
In practice, because of the inherent symmetry in the equations~\cite{bot23}, i.e. $m_x \rightarrow -m_x$, $m_y \rightarrow m_y$, $m_z \rightarrow -m_z$ (or in polar coordinates $\theta \rightarrow \pi - \theta$, $\phi \rightarrow \pi - \phi$), we need only use the $10 \ 001$ ICs which lie in the $z \ge 0$ hemisphere. We see in Fig.~\ref{fig3} that the dependence of the average voltage on the ICs corresponds exactly to the different periodicities $\mathbf{m}(t)$ in relation to the drive, i.e. many drive cycles correspond to one cycle of $\mathbf{m}(t)$. The values of the average voltages in Fig.~\ref{fig3} (b) are equal to the voltages seen along the CVC shown in Figs.~\ref{fig2} (a) and (b), for $i_{\textrm{dc}} = 1.44$ ($1.20$ and $0.95$, respectively).

We have also found that the magnetization dynamics may posses multiple stable attractors at certain values of the dc-bias, leading to additional branches in the CVCs. Such a multistability for the magnetic dynamics has been reported  in many works (see, for example,~\cite{wang2016nonlinear,wang2018emergence,contreras2022voltage,nashaat2022bifurcation,velez2020periodicity}). As we mentioned previously, for the type (ii) systems we have the unidirectional coupling between the magnetization and phase dynamics which allows us to investigate its magnetization modes separately from the phase dynamics. Using this property, we calculate the basins of attraction for the magnetization dynamics and identify 3 possible regimes at $i_{\textrm{dc}} = 1.25$, with periods 1, 2, and 4. Then we chose initial magnetization directions corresponding to different periods of the magnetization dynamics at $i_{\textrm{dc}} = 1.25$. With the help of these directions and $\varphi = 0$ as ICs, we calculate CVCs sweeping the dc-bias in up and down directions. We find that period-4 motion of the magnetization dynamics gives a rise to an additional branch shown in Fig.~4(a). In Figs.~\ref{fig4} (a-c), we show the additional branches in CVCs, together with the corresponding basins of attraction. Figs.~\ref{fig4} (b) and (c) are similar to Figs.~\ref{fig3} (a) and (b), except that there are now three possible voltage states corresponding to three different magnetic modes, respectively. In Fig.~\ref{fig3}, there were only two. Unfortunately, as we have shown in Section~\ref{sec:Stability}, not all of the detected branches are robust to the effects of thermal fluctuations. In particular, the dashed red and blue branches, shown in Fig.~\ref{fig4}(a), which are associated with the period-4 magnetization dynamics, are completely degraded by relatively small levels of noise. On the other hand, the two main branches are found to be fairly robust.

\begin{figure}[htbp]
    \centering
    \includegraphics[width=0.48\textwidth]{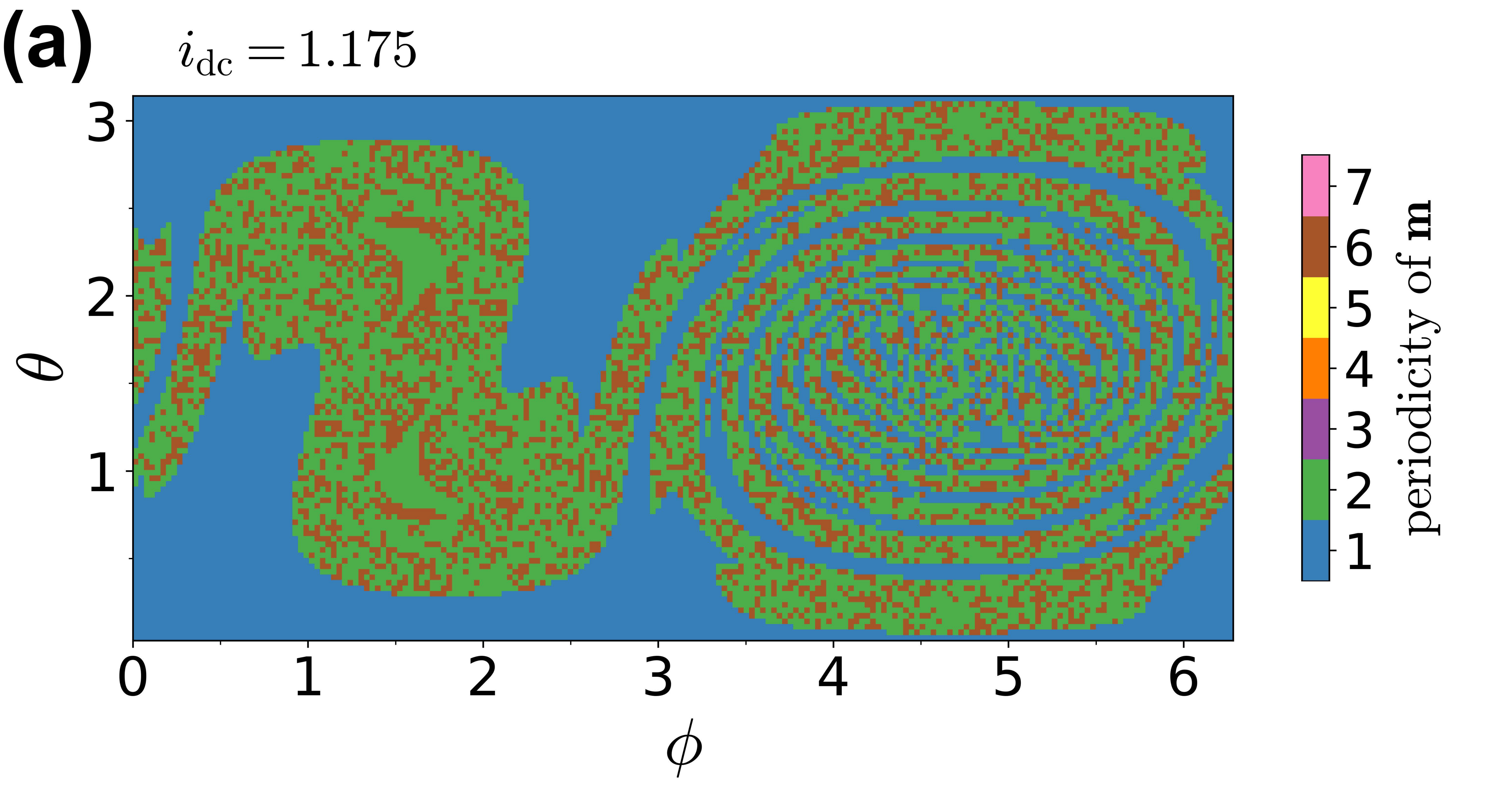}
    \includegraphics[width=0.48\textwidth]{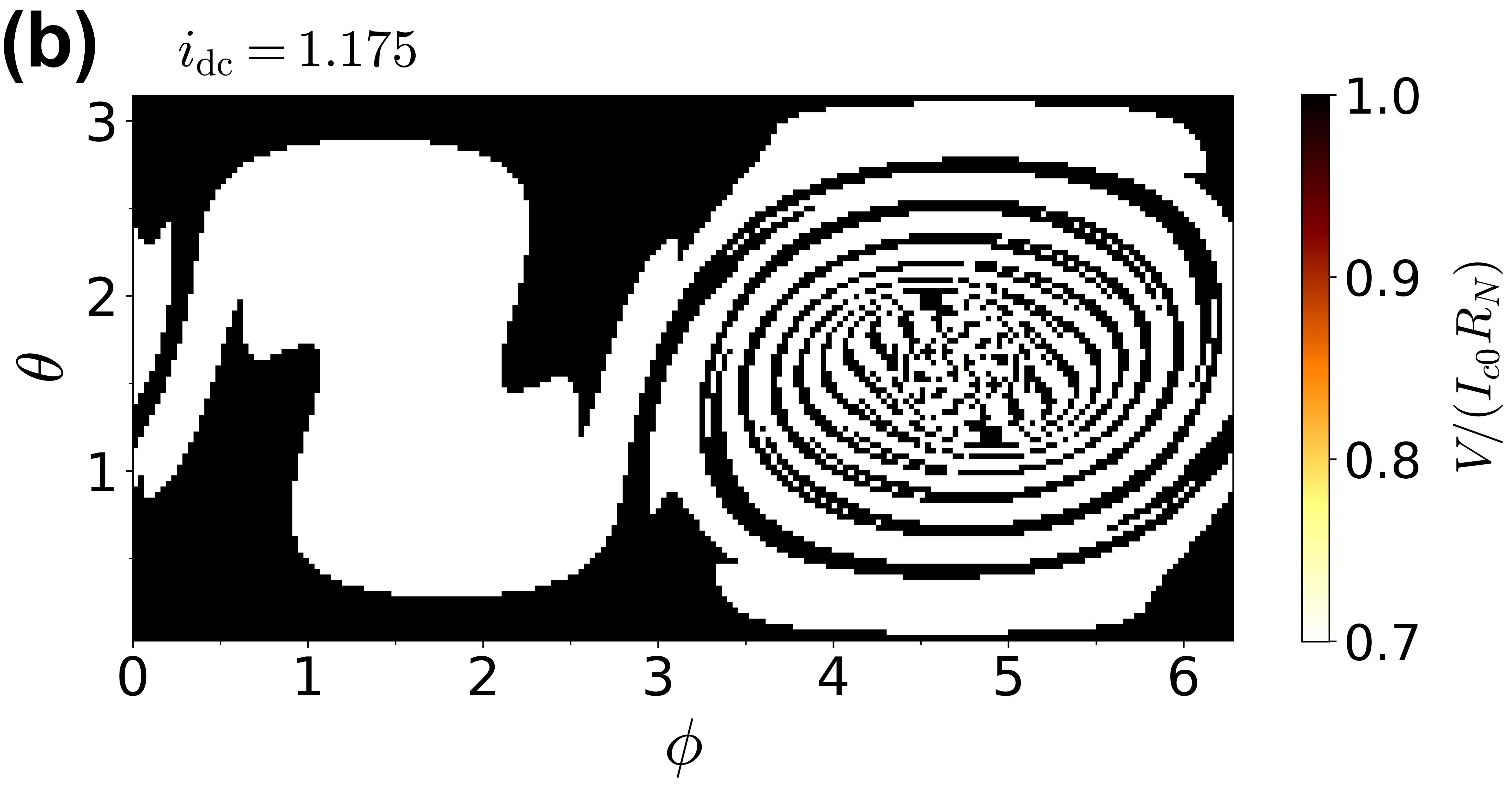}
    \caption{Broken one-to-one correspondence between the $\mathbf{m}(t)$ periodicities and $V/(I_{c0}R_N)$ at $i_{\textrm{dc}} = 1.175$, for the type (ii) system. In (a) we see that the region of period 2 behaviour (green) is interspersed by period 6 behaviour (brown), both corresponding to the Shapiro step, $V/(I_{c0}R_N) = 0.7$, represented by the white region(s) in (b). Other parameters are the same as in Figs.~\ref{fig2} to \ref{fig4}.}
    \label{fig5}
\end{figure}
The one-to-one correspondence between the regions of particular $\mathbf{m}(t)$ periodicities and average voltage does not hold for lower dc-biases. This phenomenon is shown in Fig.~\ref{fig5}, for $i_{\textrm{dc}} = 1.175$.  
We see that the period-$6$ behaviour of $\mathbf{m}(t)$, show by the brown color in (a), does not map onto a unique value for ${V}(i_{\textrm{dc}})/(I_{c0}R_N)$, as shown in (b). Moreover, at certain parameters, the boundaries between basins of attraction to different magnetization modes may become fragmented. Such fractal boundaries can in fact be seen between the period-2 (green) and period-6 (brown) basins in Fig.~\ref{fig5} (a). These appear to form a so-called riddled basin~\cite{ale92,caz01}, for which every point in one basin is arbitrarily close to some point in another. Consequently, it becomes extremely difficult to discover {\em all} possible branches in the CVCs of our system, and we do not claim to have done so here.

A detailed analysis of the system is provided in the Appendix. It shows that the breaking of the one-to-one correspondence depends on the relative amplitudes of the Fourier harmonics in the frequency spectrum of $i_{c}(m_x)$.

\subsection{Type (iii) systems}
In the third system type, the magnetization of the interlayer depends explicitly on the phase difference $\Phi = \varphi - r m_y$ (see, Table~\ref{tab:data}). As a result of this dependence, the dc-bias alone produces a time-varying torque on the magnetization and one may therefore expect to see interesting dynamical features, including chaos, in CVCs of the type (iii) system without ac driving.

\begin{figure}[hbtp!]
    \centering
    \includegraphics[width=0.48\textwidth]{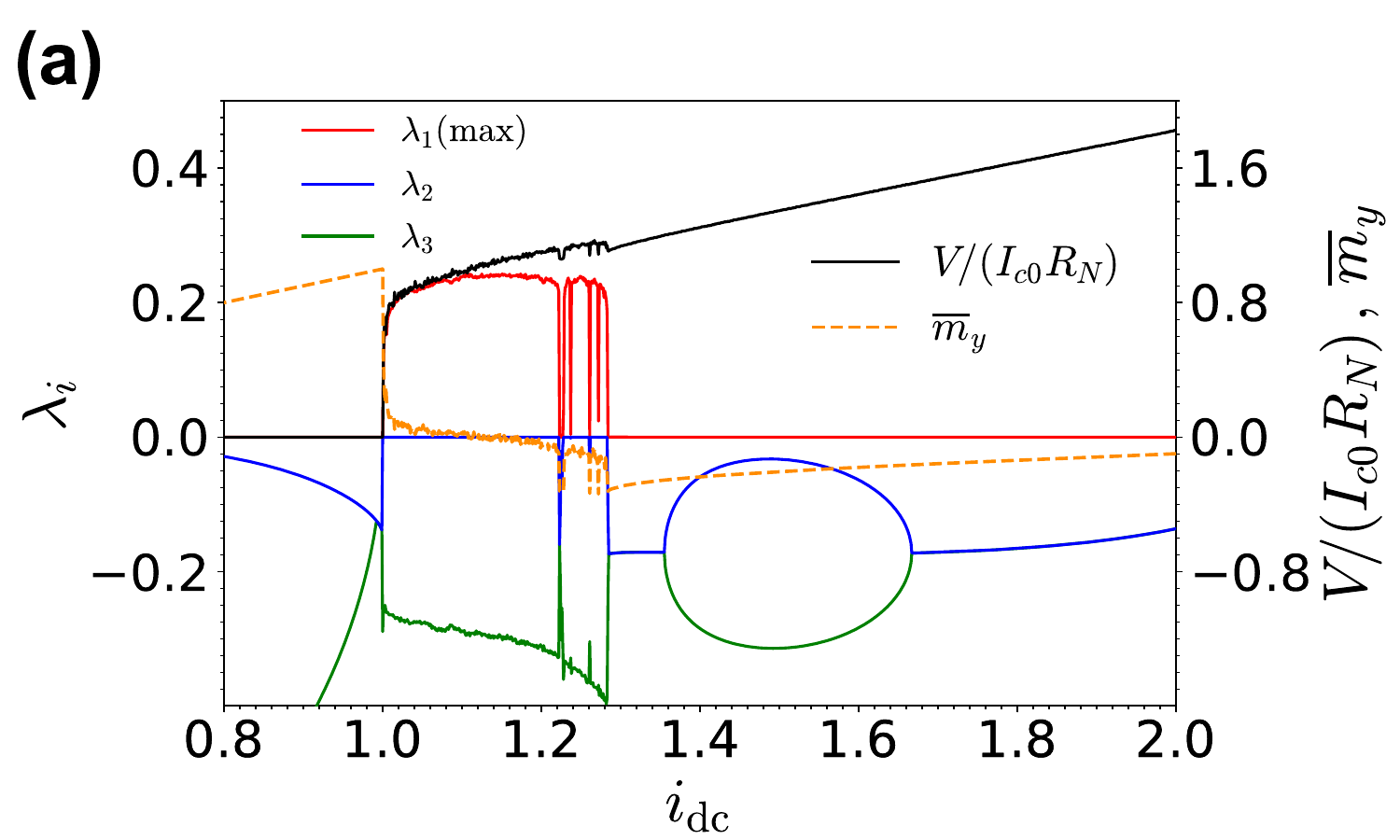}
    \includegraphics[width=0.48\textwidth]{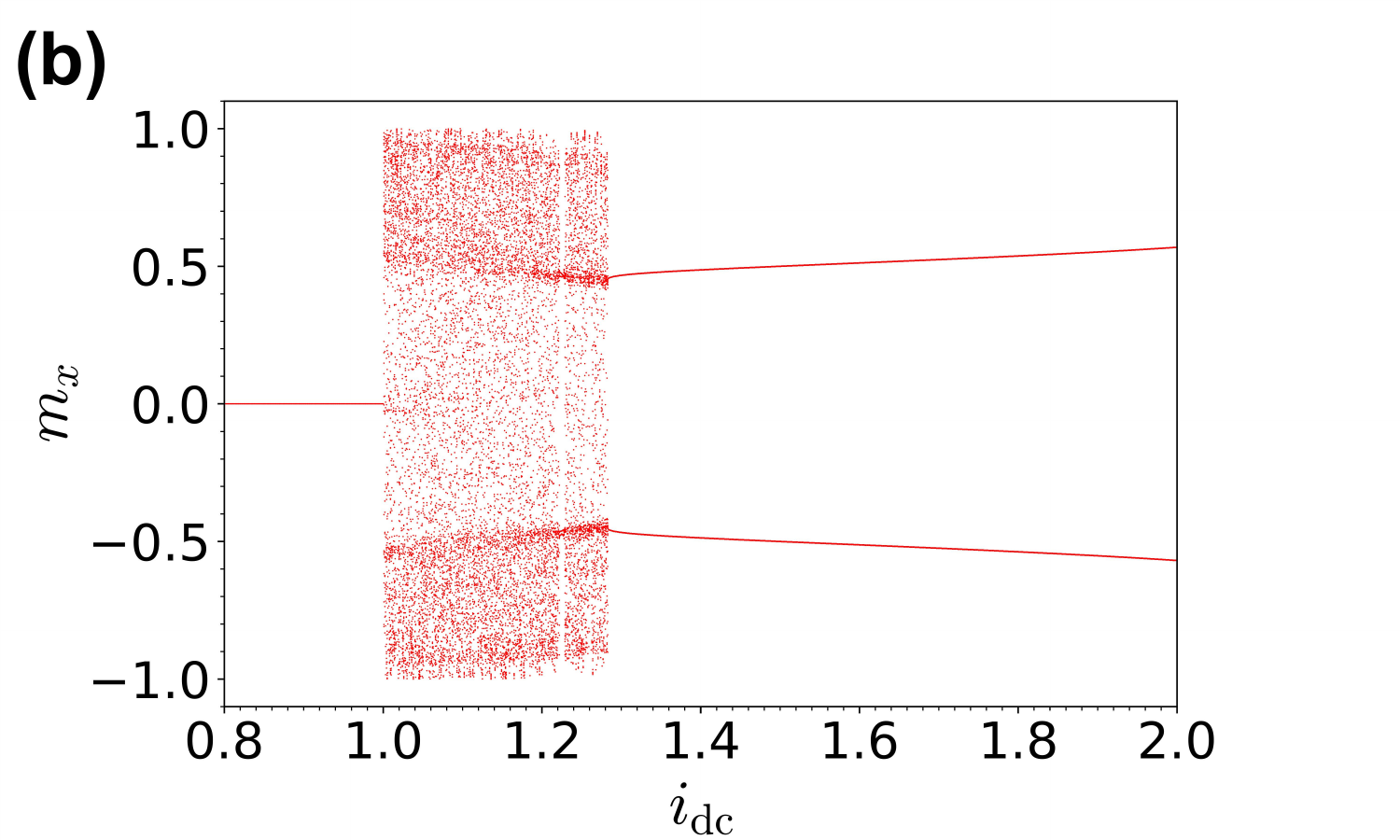}
    \includegraphics[width=0.48\textwidth]{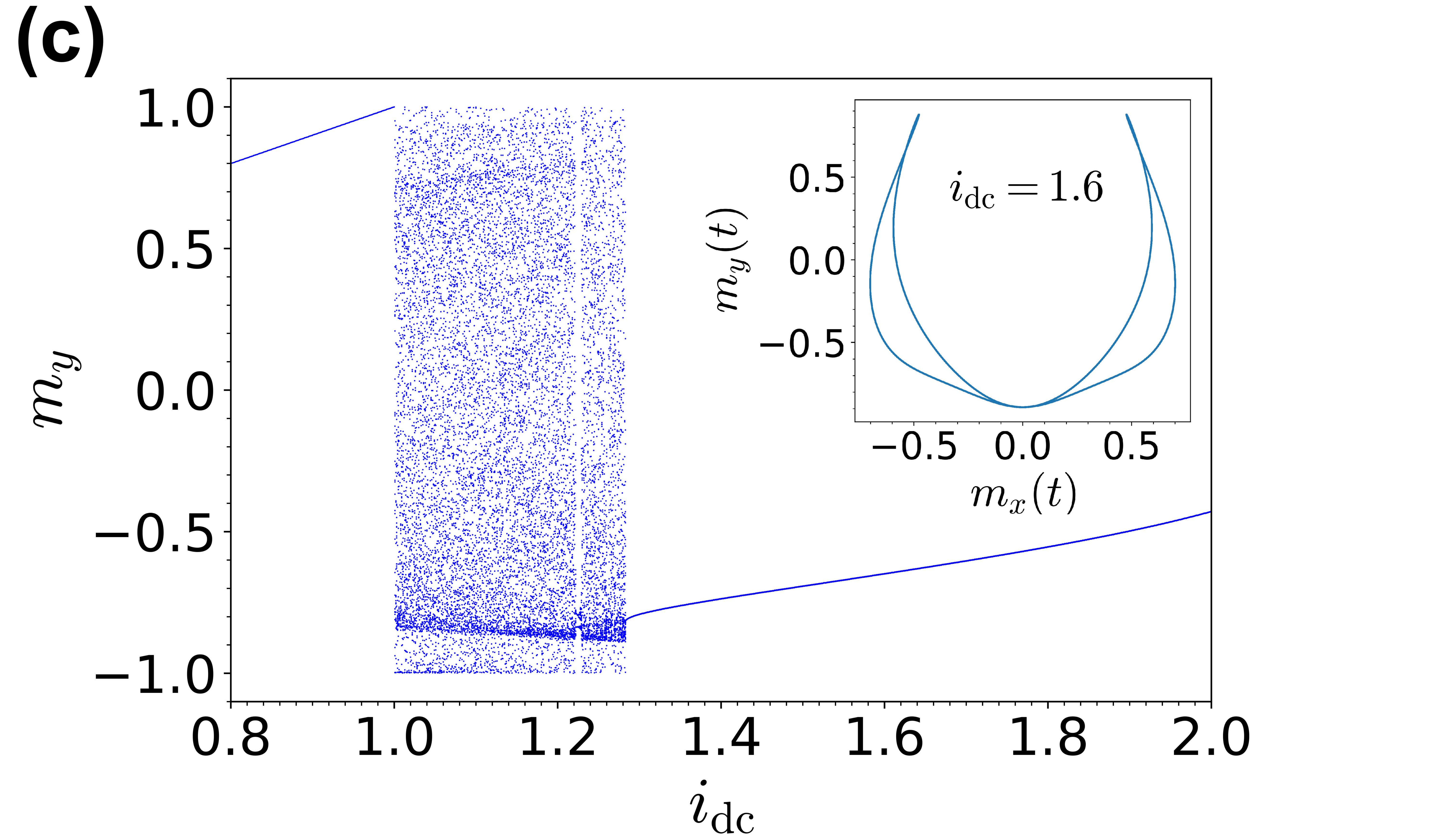}
    \caption{Dynamical response of the type (iii) system to sweeps in the dc-bais current, without the ac drive. In this case, the up and down sweep directions produce the same response. (a) Lyapunov exponents ($\lambda_i$, $i = 1,2,3$), time averaged voltage $V/(I_{c0}R_N)$ and time averaged $\overline{m}_y$. In (b) and (c) are shown the orbit diagrams for $m_x$ and $m_y$, respectively. The values shown along the vertical axes are plotted as points at the instant when $\sin(\varphi-rm_y)$ crosses from negative to positive. The inset in (c) shows the inherent symmetry in the typical periodic motion that occurs for $i_{\mathrm{dc}} \gtrapprox 1.284$, i.e. to the right of the chaotic region. See main text for details. Parameters: $i_{\textrm{ac}} = 0$, $G = 0.2$, $r = 5$, $\alpha = 0.02$, $w = 1.0$, and $\tilde{d}_F =1$.} \label{fig6}
\end{figure}

\begin{figure*}[hbtp]
    \includegraphics[width=0.48\textwidth]{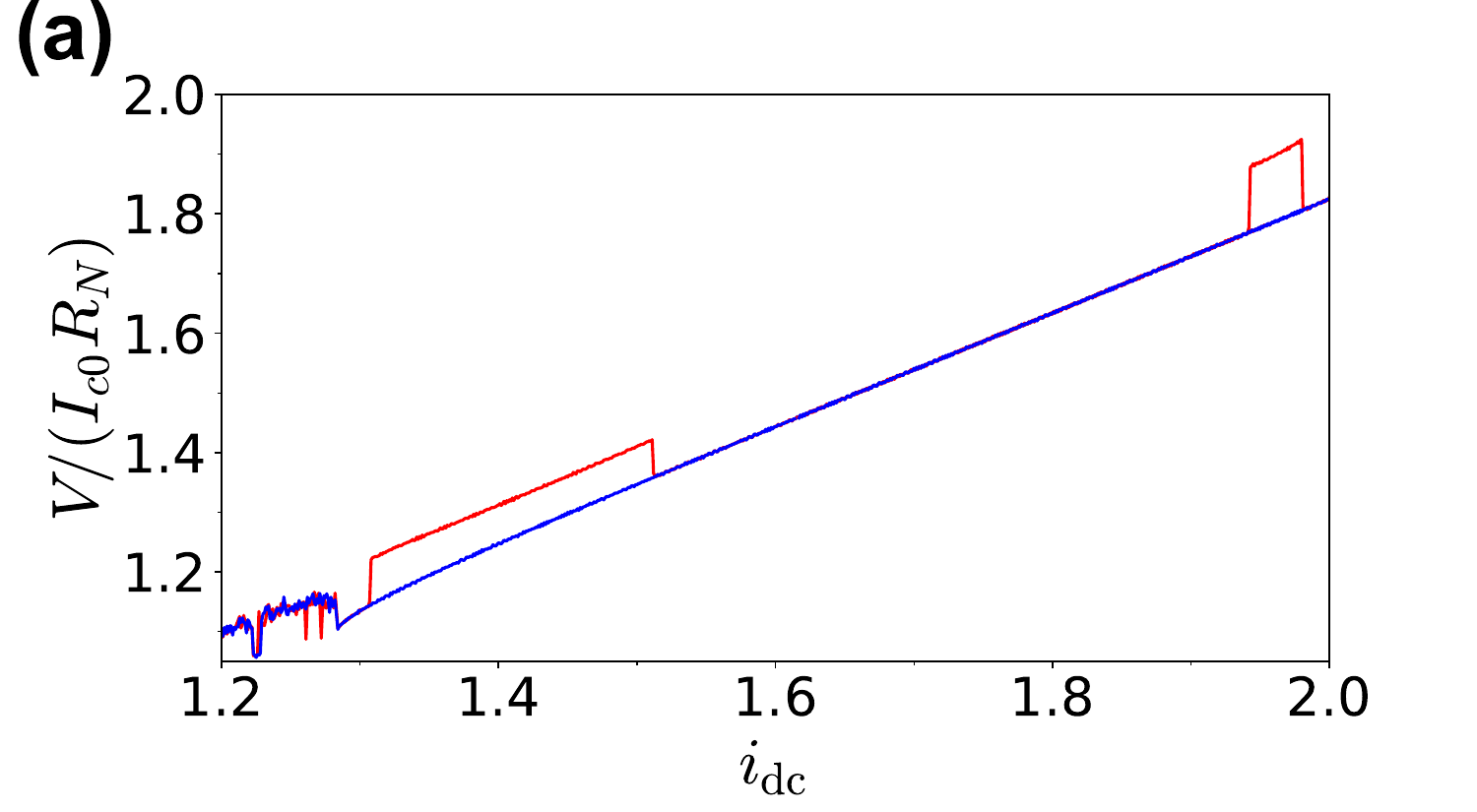}
    \includegraphics[width=0.48\textwidth]{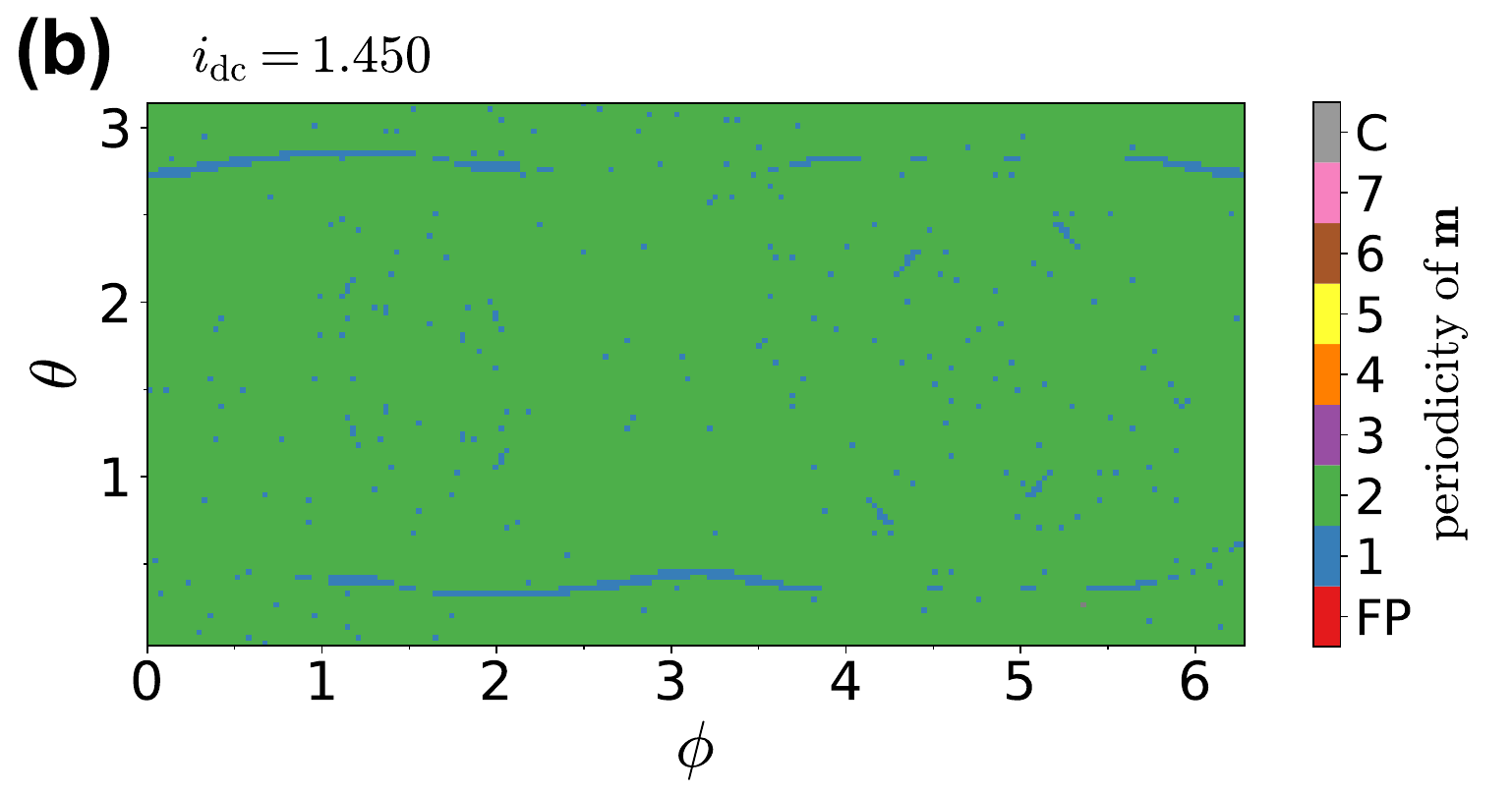}
    \includegraphics[width=0.48\textwidth]{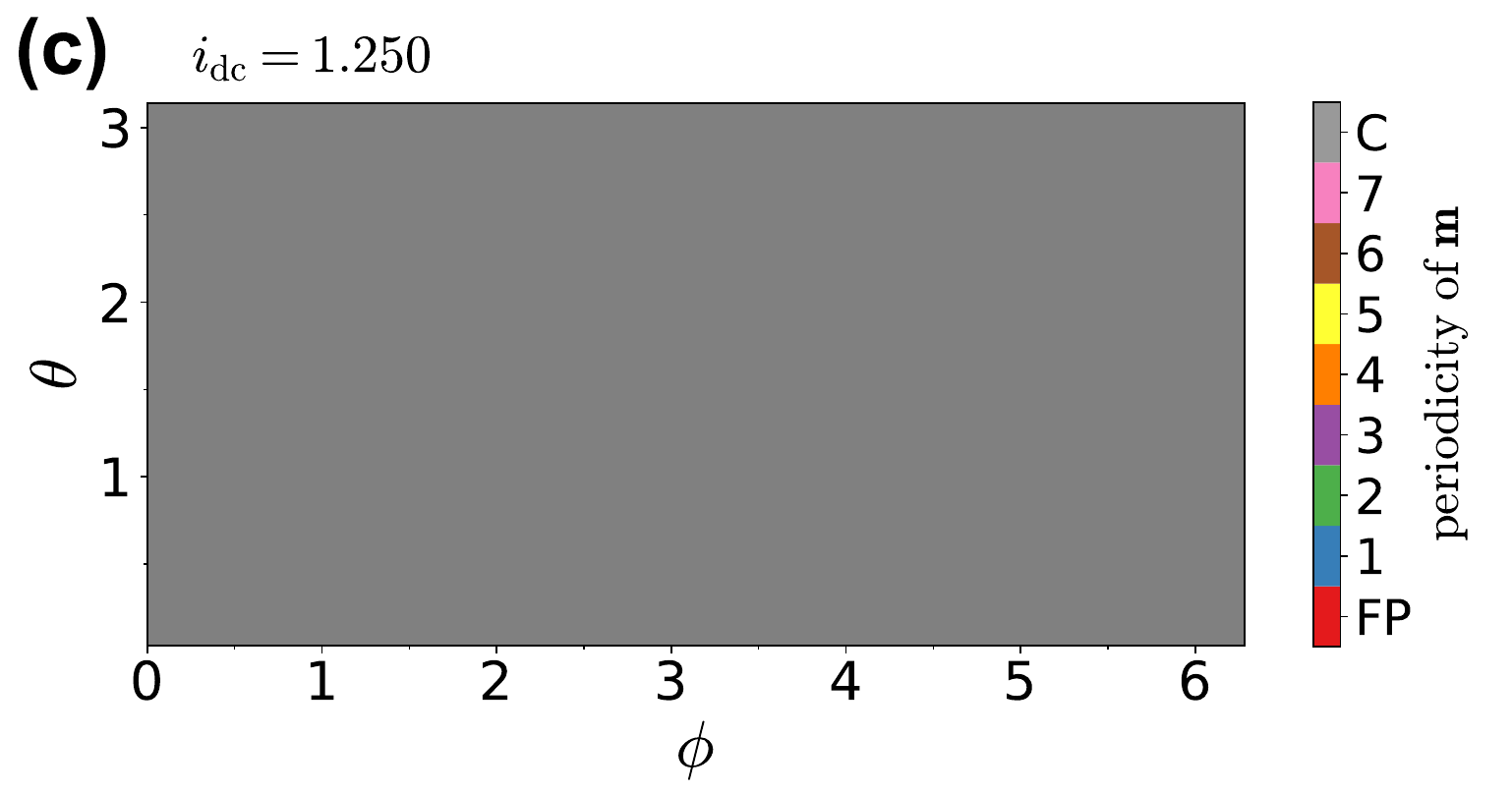}
    \includegraphics[width=0.48\textwidth]{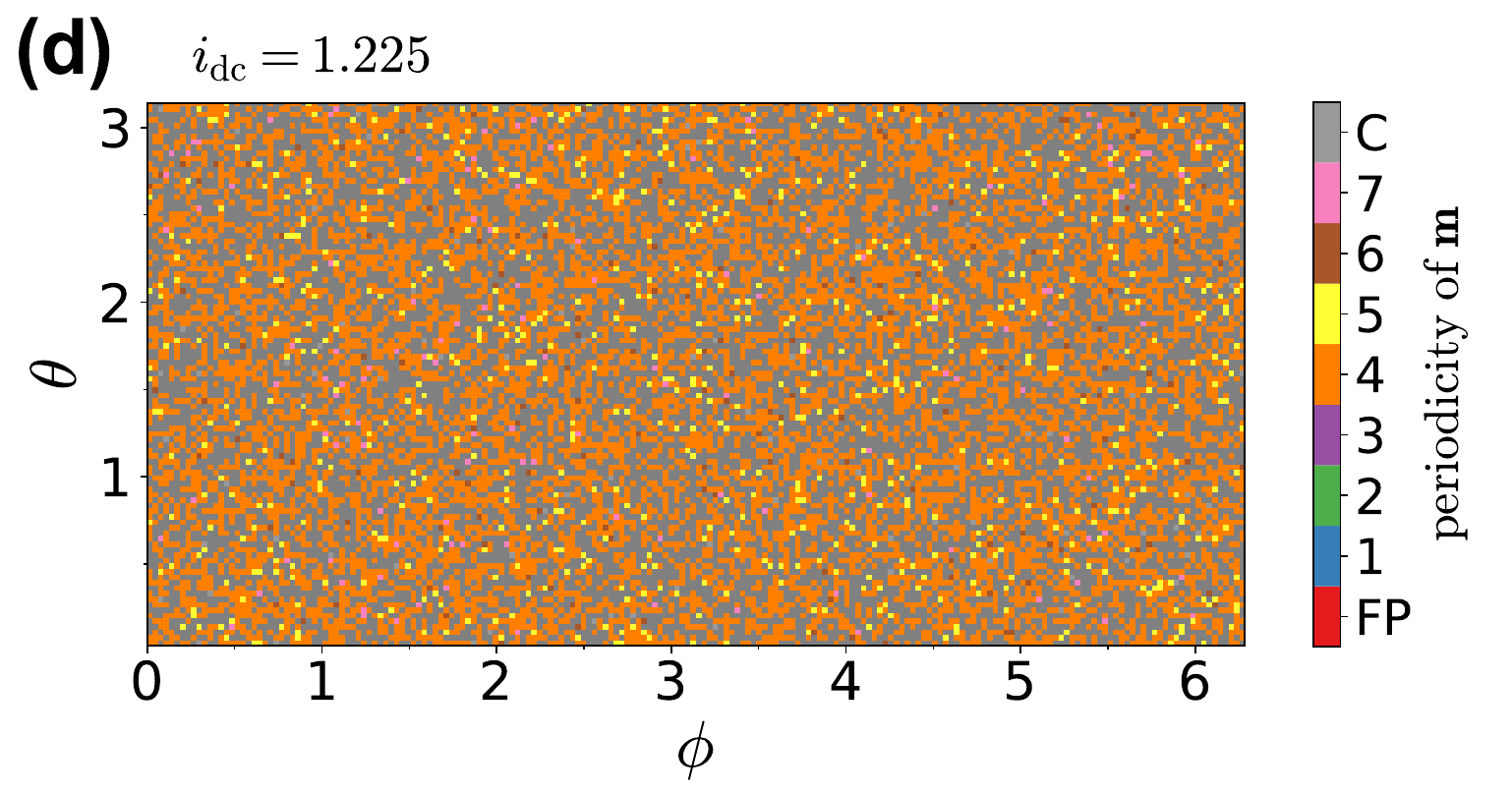}
    \caption{Time averaged voltage (a) and the basins of attraction at three different dc-biases (b-d), for the type (iii) system. (a) The two different branches in CVC which are in one-to-one correspondence with period 1 (red curve) and period 2 (blue curve) behaviour in the magnetic subsystem. The CVC seen previously in Fig.~\ref{fig6} (a) coincides with the blue curve for period 2. (b) Coexistence of period 1 and period 2 behaviour in the basin of attraction at $i_{\mathrm{dc}} = 1.450$. (c) Chaotic regime at $i_{\mathrm{dc}} = 1.250$. (d) Coexistence of chaos and periods 5, 6, and 7 behaviour within the periodic window surrounding $i_{\mathrm{dc}} = 1.225$. Other parameters are as in Fig.~\ref{fig6}. In the color scales on the right, FP and C stand for fixed point and chaos, respectively. }
    \label{fig7}
\end{figure*}

As an example, we show in Fig.~\ref{fig6} the results for the type (iii) system with $w=1$ and $i_{\textrm{ac}} = 0$. Without the ac drive, there are no Shapiro steps in the CVC, as shown in Fig.~\ref{fig6} (a). Since the effective dimension of the state space is only three, i.e. two dimensions for the magnetic subsystem and only one for that of the overdamped JJ, the Lyapunov exponent spectrum consists of only three exponents, one being trivially zero, as usual. As the dc-bias is varied, we encounter three distinct regions: {\em periodic} -- for $i_{\mathrm{dc}} \gtrapprox 1.284$, {\em chaotic} -- for $1 < i_{\mathrm{dc}}  \lessapprox 1.284$ and a {\em fixed point} -- for $i_{\mathrm{dc}} \le 1$. Because there is no external ac drive here, we measure the periodicity of the system relative to the oscillations of $\sin(\varphi-rm_y)$. Specifically, in our numerical simulations we detect how many oscillations of $\sin(\varphi-rm_y)$ correspond to one cycle of $\mathbf{m}(t)$, as explained in~\cite{bot23}. Comparing Figs.~\ref{fig6} (b) and (c) we see the periodicities of $m_x$ and $m_y$ differ in the $i_{\mathrm{dc}} \gtrapprox 1.284$ region. We find that the periodicities of $m_x$ and $m_z$ (or $m_y$, $\varphi$ and $\dot{\varphi}$) are always the same in this system, which is why orbits diagrams for $m_z$, $\varphi$ and $\dot{\varphi}$ have been omitted in Fig.~\ref{fig6}. The period-2 and period-1 behaviour of $m_x$ and $m_y$, respectively, is related to the inherent symmetry of the equations, which can be seen clearly in the projection of the trajectory, as shown by the inset of Fig.~\ref{fig6} (c). The onset to fully developed chaos occurs abruptly at $i_{\mathrm{dc}} \approx 1.284$, without any transitioning through intermediatory quasi-periodic or period doubling regions, as would usually be the case for an ordinary JJ. Finally, for $i_{\mathrm{dc}} \le 1$, the dynamics ceases as the system is attracted to one of the fixed points: $m_x = 0$, $m_y = Gr i_{\mathrm{dc}} =  Gr \sin(\varphi - r m_y)$, and $m_z = \pm\sqrt{1 - m^2_y}$. In the orbit diagrams, (b) and (c), when $\sin(\varphi - r m_y)$ does not oscillate, we plot these equilibrium values of $m_x$ and $m_y$. 

We have also generated additional branches in the CVCs and basins of attraction for the type (iii) system. In Fig.~\ref{fig7} (a) we have used different ICs to generate the two main branches in the CVCs, corresponding to period 1 and 2 behaviours in the magnetization. The corresponding periodicities of the magnetization can be seen in Figs.~\ref{fig7} (b-d), which show the basins of attraction at three selected dc-biases. In Fig.~\ref{fig7} (b) the bias current falls within the periodic region and we see that the magnetization vector $\mathbf{m}$ may exhibit either period-1 or period-2 behavior, depending on the ICs for $\theta$ and $\phi$. The period two behaviour here corresponds to orbit diagrams seen previously in Fig.~\ref{fig6} (b) and (c), where we saw that $m_x$ and $m_y$ had the periodicities 2 and 1, respectively. Obviously, the magnetization vector has the periodicity corresponding to the lowest common multiple of the periodicities for its components. On the other hand, the period-1 behavior in $\mathbf{m}$ is a different regime, which gives rise to a different average voltage to that shown in Fig.~\ref{fig6} (a). Thus, even for $i_{\mathrm{ac}} = 0$, there occur different magnetization states which in turn can produce multiple branch structure in the CVCs. Here, since there is always a one-to-one correspondence between the magnetization modes and the average voltages within the periodic region, we have not show the corresponding basins for the average voltage. At other parameters, however, the one-to-one correspondence could be broken, as we saw previously for the type (ii) system. In Fig.~\ref{fig7} (c) we see that, at $i_{\mathrm{dc}} = 1.250$, the chaos is global, i.e. it does not depend on the ICs. On the other hand, within the chaotic region there are windows of periodic behavior which arise only for certain ICs; such as, the one that can be seen in Fig.~\ref{fig6} (a) at $i_{\mathrm{dc}} = 1.225$. As shown in Fig.~\ref{fig7} (d), at $i_{\mathrm{dc}} = 1.225$, the basin of attraction within the periodic window contains mostly period-4 behavior, but it is also shared to lesser degrees by periods-5, 6, 7 and chaotic behaviour (corresponding to grey in the color bar). Moreover, as we have mentioned before, the basins are not separated by clear boundaries, but rather, are highly fragmented.

\section{Dynamical switching of the JJ voltage states via a current pulse}
\label{sec:Switching}
In this section we demonstrate that it is possible, without capacitance, to switch between different dynamical voltage states, which are seen in the previous sections. We express the switching current pulse in terms of two Heaviside step functions, $\theta (t)$, as 
\begin{equation}\label{eq8}
    i_{p}(t) = A_s \left[\theta(t-t_0) - \theta(t - \delta t - t_0)\right]\sin\left[\Omega(t - t_0)\right],
\end{equation}
where $A_s$ is the height of the pulse, $\delta t$ is its width, and $t_0$ is its start time. To apply the switch in our models we simply add $i_{p}(t)$ to the total current, $i_{\mathrm{dc}} + i_{\textrm{ac}}\sin \left( \Omega t\right)$ in Eqs.~(\ref{eq6}), after the usual transient time.

For simplicity, we discuss only the type (ii) model at the dc-bias, $i_{dc} = 1.44$. At this current there exist only two magnetization regimes (see, Fig.~\ref{fig3}): the fixed points $\mathbf{m} = (0,\ \pm1,\ 0)$, and the period-1 motion. These regimes correspond to the average voltages $V/(I_{c0}R_N) = 0.95$ and $V/(I_{c0}R_N) = 1.20$, respectively (see, Figs.~\ref{fig2} (a) and (b)). 
\begin{figure}[t]
    \centering
    \includegraphics[width=0.48\textwidth]{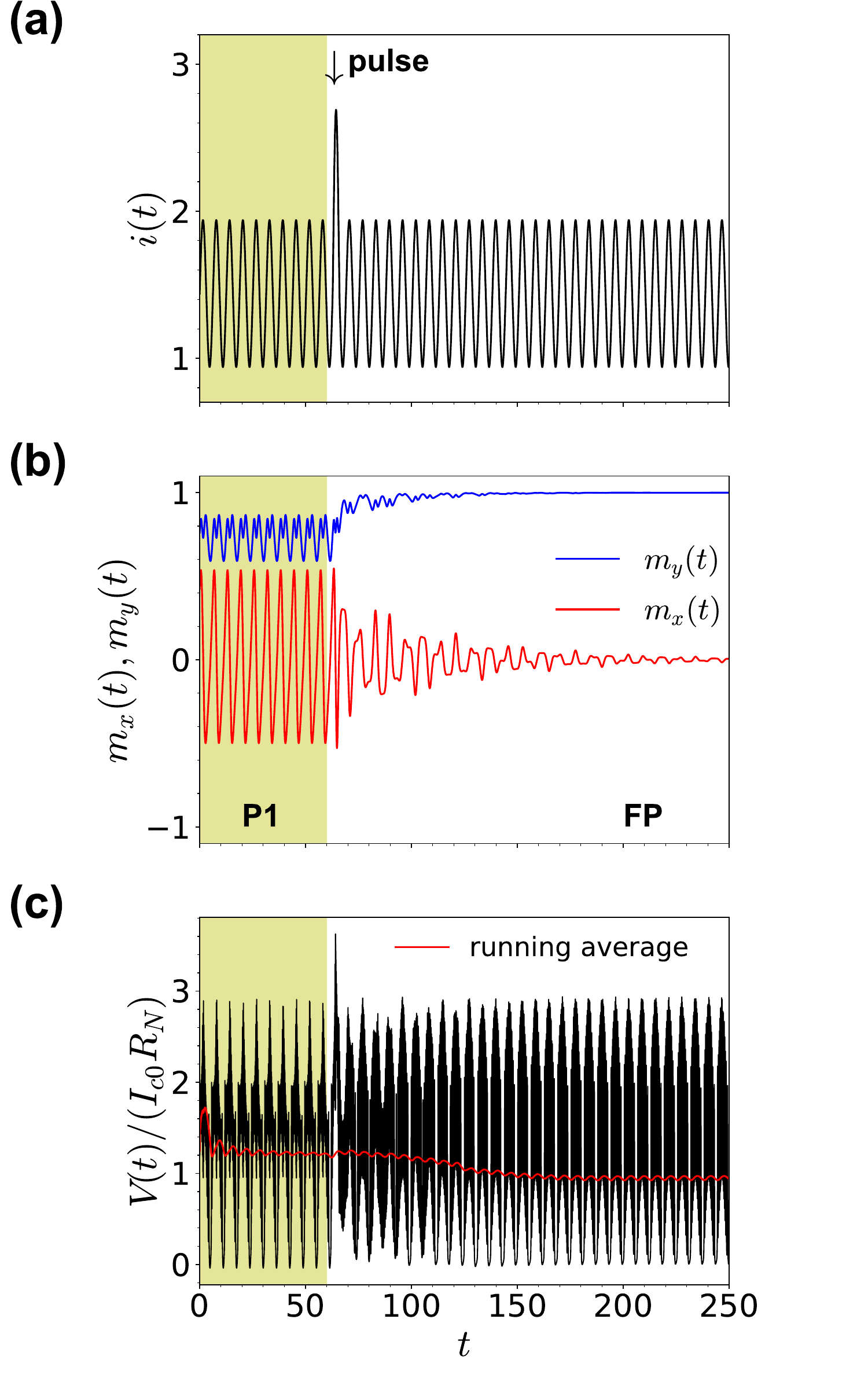}
    \caption{Magnetization dynamics in the type (ii) system under the current pulse. (a) The applied current pulse is given by Eq.~(\ref{eq8}) and is shown by the black arrow. The parameters for the pulse are $A_s = 0.75$, $\delta t = \tau/2$, and $t_0 = 10\tau$. The initial condition used here for the period-1 behaviour is $\theta=\phi=2$ and $\varphi=0$, cf. Fig.~3(a). Other parameters are the same as in Figs.~\ref{fig2} and~\ref{fig3}, i.e. $i_{\textrm{dc}} = 1.44$, $i_{\mathrm{ac}} = 0.5$. (b) The switching from period-1 motion (P1 - shaded regions) with $V/(I_{c0}R_N) = 1.20$ to the fixed point (FP) $\mathbf{m} = (0,\ 1,\ 0)$ with $V/(I_{c0}R_N) = 0.95$. (c) Time series of the change in voltage that occurs as a result of the switching. The running average (dashed red line) has been computed over 50 time units, using 5000 samples.}
    \label{fig8}
\end{figure}
As shown in Fig.~\ref{fig8}, the application of the pulse switches the system from the upper branch ($V/(I_{c0}R_N) = 1.20$) to the lower branch ($V/(I_{c0}R_N) = 0.95$).

In our present simulations, it was not possible to switch back from the state with $V/(I_{c0}R_N) = 0.95$ at $i_{\textrm{dc}} = 1.44$ to the state $V/(I_{c0}R_N) = 1.20$ at $i_{\textrm{dc}} = 1.44$, due to the fact that, when the magnetization is located at one of the fixed points $\mathbf{m} = (0,\ \pm1,\ 0)$, the torque produced by the current pulse cannot move the magnetization away from the fixed point, even if the fixed point becomes unstable (because $\mathbf{T}_2$ always acts parallel to $\mathbf{m}$). Thus, in order to achieve switching in both directions, we need to add thermal noise in the form of the thermal field~\cite{Guarcello20}, rather than in the form of a thermal current~\cite{guarcello2021thermal}. The thermal noise directly affects the magnetization dynamics, thereby allowing the system to switch back out of the fixed points, provided that they are unstable, of course. However, a full discussion of the switching properties of the three types of system present, such as recently done by Guarcello et al.~\cite{guarcello2023switching} for the case of an underdamped junction, is beyond the scope of the present work.

In the next section, we will briefly show how the obtained features of the CVCs behave under the influence of thermal fluctuations.

\section{Stability of the JJ voltage states to thermal fluctuations} \label{sec:Stability}

Until now there have been relatively few studies that have included the effects of noise in systems of Josephson junctions coupled to the LLG equation~\cite{Guarcello20,guarcello2021thermal,guarcello2023switching}. However, the effects of noise on the two separate subsystem (JJ and LLG), have been studied intensively. For example, the magnetization dynamics with thermal noise has been studied in Refs.~\cite{leliaert2017adaptively,nis15,nis18,rom14}. Also the Josephson junction dynamics in the presence of thermal noise has been investigated in Refs~\cite{barone1982physics,Kautz90noise,BLACKBURN20161,GUARCELLO2021111531,Filatrella02,tafuri2019fundamentals,hansen1991noise}. In the present section we investigate the coupled dynamics of the JJ and its interlayer magnetization in the presence of thermal noise.

In the previous sections we have seen that the CVCs of overdamped $\varphi_0$ Josephson junctions exhibit some unusual features that are related to the induced magnetization dynamics in the interlayer. These features included chaotic states and, in particular, hysteresis that led to multiple branches in the CVCs. Since the latter feature could facilitate certain cryogenic switching applications, it is important to test how it may be affected by thermal noise, which is inevitably present in real systems. Therefore, in this section,  we will calculate the noise-averaged CVCs corresponding to the branches found earlier. As we shall see, for typical noise levels, we may still expect to achieve switching, such as we have seen in connection with~Fig.~\ref{fig8}, without noise. Furthermore, while in Fig.~\ref{fig8} we were unable to switch the system back from the FP to the period-1 behavior, we find that the thermal noise sufficiently destabilizes the fixed point to allow switching in both directions.

As a general study of the influence of thermal noise on $\varphi_0$ JJs is beyond the scope of our present work, we focus here again on type (ii) systems with the same parameters as before: $G = 0.2$, $r = 5$, $\alpha = 0.02$, $\Omega = 1$, $i_{\mathrm{ac}} = 0.5$, $w = 0.05$, and $\tilde{d}_F = 1$. Following the latest methods employed in other studies involving noise in magnetic systems  ~\cite{Guarcello20,guarcello2021thermal,guarcello2023switching}, we add the thermal noise to Eq.~(\ref{eq6}), as follows: 
\begin{subequations} 
\label{eq7}
\begin{eqnarray}
\dot{\mathbf{m}} & = &  - \frac{1}{1+\alpha^2}\left\{\mathbf{m}\times
\left( \mathbf{T}_{2}+m_{z}\mathbf{\hat{e}}_{z} + \mathbf{h}_{\textrm{th}}\right) + \right.   \nonumber \\ 
& & \left.   +\alpha \mathbf{m\times }\left[ \mathbf{m}\times \left( \mathbf{T
}_{2}+m_{z}\mathbf{\hat{e}}_{z} + \mathbf{h}_{\textrm{th}}\right) \right]\right\},  \label{eq7a}\\
\mathbf{T}_2 &=& Gr \left(i_{\textrm{dc}} + i_{\textrm{ac}}\sin\left(\Omega t\right) + i_{\textrm{th}}\right)\hat{\mathbf{e}}_y,\label{eq7b} \\
w\dot{\Phi} & = &   i_{\textrm{dc}}  + i_{\textrm{ac}}\sin \left( \Omega
t\right) + i_{\textrm{th}} - i_{c}\left( m_{x}\right) \sin \Phi .   \label{eq7c}
\end{eqnarray}
\end{subequations}
Here, $\mathbf{h}_{\textrm{th}}$ and $i_{\textrm{th}}$ represent the fluctuating thermal field and the noise in current, respectively. The noise sources obey the white noise correlation properties:
\begin{equation}
\begin{aligned}
    &\langle i_{\textrm{th}}(t) \rangle = 0, \, \,  \langle i_{\textrm{th}}(t) i_{\textrm{th}} (t') \rangle = 2 D_I \delta(t-t'), \\
    & \langle h_{i,\textrm{th}}(t)\rangle = 0, \,  \, \langle h_{i,\textrm{th}}(t) h_{j,\textrm{th}} (t') \rangle = 2D_H\delta_{ij}\delta(t-t').
\end{aligned} \label{eq9}
\end{equation}
In Eq.~(\ref{eq9}), the notation $\langle ... \rangle$ denotes the average over noise realizations and $i,j \equiv x,y,z$. The dimensionless coefficients are given by $D_I = w k_B T/E_J$ and $ D_H = \alpha k_B T/(K \mathcal{V}) = \alpha G D_I/w$, where $T$ is the temperature. $D_I$ and $D_H$ determine the respective noise intensities in the system. For $T_c \approx 10\ K$, $I_{c0} \approx 0.45\ \mu A$, $T \approx T_c/10$, $E_J = \frac{\hbar I_{c0}}{2 e}$ we estimate $D_I \sim 5\times 10^{-3}$. Using the further variation of $T$, one may gain additional control over $D_I$  that may take values from $D_I \sim  5 \times 10^{-4} \text{ to } 5 \times10^{-2}$.  For low temperatures $T \ll T_c$, we estimate the relation between $D_H$ and $D_I$ to be $D_H \approx  0.08 D_I \ll D_I$.

The noise induced by $i_{\textrm{th}}$ and $h_{i,\textrm{th}}$ enters (\ref{eq7}) multiplicatively, that is why we use the Stratonovich prescription for the stochastic integration (see, for example,~\cite{rom14}). This numerical scheme treats the magnetization in spherical polar coordinates and is given by:
\begin{widetext}
\begin{equation} \label{eq10}
\begin{aligned}
    \Phi^{n} - \Phi^{n-1} & = \frac{1}{w}\left[i_{\textrm{dc}} + i_{\textrm{ac}} \sin \Omega t^{n-1} - i_c\left(\frac{m^{n}_x + m^{n-1}_x}{2}\right)\sin\left(\frac{\Phi^{n} + \Phi^{n-1}}{2}\right)  \right]\Delta t + \frac{\Delta i_{\textrm{th}}}{w};\\
    h^{n-1}_\theta & = G r(i_{\textrm{dc}} + i_{\textrm{ac}}\sin\Omega t^{n-1}) \cos\left(\frac{\theta^n + \theta^{n-1}}{2}\right)\sin\left(\frac{\phi^n + \phi^{n-1}}{2}\right) - \frac{1}{2} \sin\left(\theta^n + \theta^{n-1}\right);\\
    h^{n-1}_\phi & =  G r(i_{\textrm{dc}} + i_{\textrm{ac}}\sin\Omega t^{n-1}) \cos\left(\frac{\phi^n + \phi^{n-1}}{2}\right);\\
    \Delta h^{n-1}_\theta & = \Delta h_{x,\textrm{th}} \cos\left(\frac{\theta^n + \theta^{n-1}}{2}\right)\cos\left(\frac{\phi^n + \phi^{n-1}}{2}\right) - \Delta h_{z,\textrm{th}} \sin\left(\frac{\theta^n + \theta^{n-1}}{2}\right) \\
    & \qquad \qquad \qquad \qquad \qquad  + \left[\Delta h_{y,\textrm{th}} +  Gr  \Delta i_{\textrm{th}} \right] \cos\left(\frac{\theta^n + \theta^{n-1}}{2}\right)\sin\left(\frac{\phi^n + \phi^{n-1}}{2}\right);\\
    \Delta h^{n-1}_\phi & = \left[\Delta h_{y,\textrm{th}} +  Gr \Delta i_{\textrm{th}}\right]\cos\left(\frac{\phi^n + \phi^{n-1}}{2}\right) - \Delta h_{x,\textrm{th}} \sin\left(\frac{\phi^n + \phi^{n-1}}{2}\right);\\ 
    \theta^{n} - \theta^{n-1} & = \frac{\Delta t}{1+\alpha^2}\left[ h^{n-1}_{\phi} + \alpha   h^{n-1}_\theta  \right] + \frac{1}{1+\alpha^2}\left[
    \Delta h^{n-1}_\phi +  \alpha \Delta h^{n-1}_\theta\right] ;\\
    \phi^{n} - \phi^{n-1} & = \left\{ (1+\alpha^2)\sin\left(\frac{\theta^n + \theta^{n-1}}{2} \right)\right\}^{-1}  \left\{ \Delta t \left[- h^{n-1}_\theta + \alpha h^{n-1}_\phi   \right] -\Delta h^{n-1}_\theta + \alpha \Delta h^{n-1}_\phi  \right\}  .
\end{aligned}
\end{equation}
\end{widetext}

Here $\Delta t$ is a constant time step and $n$ is the discrete time index ($t_n = n\Delta t$) for the state variables $\left\{\Phi^{n}\right\}$, $\left\{\theta^{n}\right\}$, and $\left\{\phi^{n}\right\}$. The noise terms $\Delta i_{\textrm{th}} \sim \mathcal{N}(0, 2 D_I \Delta t)$, $\Delta h_{i\textrm{th}} \sim \mathcal{N}(0, 2D_H \Delta t)$ are normally distributed with zero mean and variances $ 2 D_I \Delta t$, $2 D_H \Delta t$, respectively.

\subsection{CVCs in the presence of thermal fluctuations} \label{subsec:deformation}
\begin{figure}[t]
\includegraphics[width=0.48\textwidth]{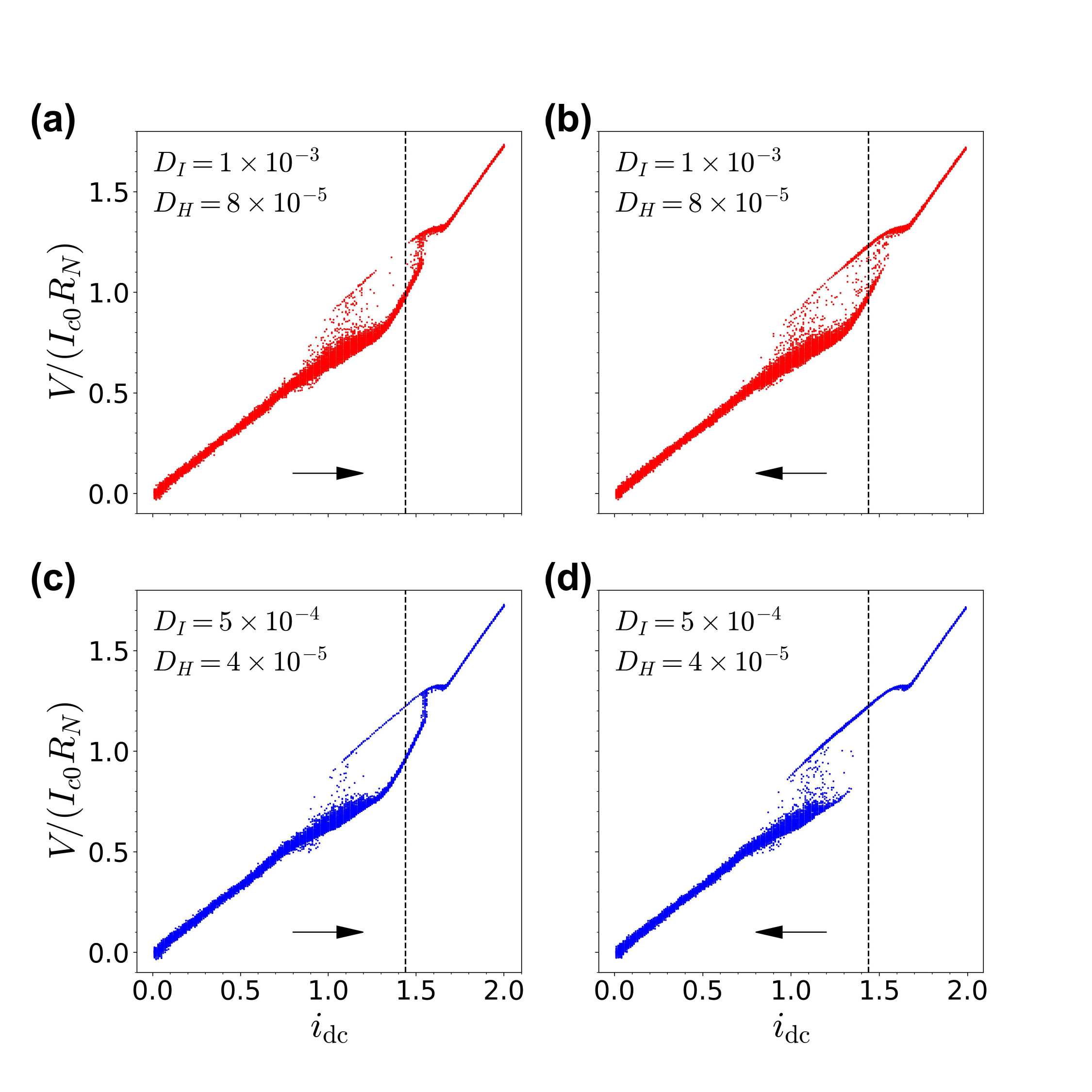}
\caption{CVCs of the type (ii) systems at two levels of the noise intensity: (a-b) $(D_I,\ D_H) = (10^{-3},\ 8\times 10^{-5})$, and (c-d) $(D_I,\ D_H) = (5\times 10^{-4},\ 4\times 10^{-5})$.  Horizontal arrows indicate the direction of the current sweep. The dashed vertical lines indicate $i_{\textrm{dc}} = 1.44$, i.e. the current corresponding to where the switching was investigated in Fig.~\ref{fig8}. Other parameters are the same as in Figs.~\ref{fig2}-\ref{fig5} and \ref{fig8}. }
\label{fig9}
\end{figure}

We have recalculated the CVCs that were seen previously in Fig.~\ref{fig2}, with the help of the stochastic Eq.~(\ref{eq10}). We used a time step $\Delta t = 0.001$, which was found to be sufficiently small to ensure proper convergence. To obtain good statistics we swept the dc bias up and down, $100$ times in each direction, and for two low-temperature noise intensities: $(D_I,\ D_H) = (1,0.08)\times 10^{-3}$ and $(D_I,\ D_H) = (5, 0.4)\times 10^{-4}$. The results are presented in Fig.~\ref{fig9}. 
As one may expect, the thermal noise tends to `wash out' the Shapiro steps (c.f. Fig.~\ref{fig2}) in both of the branches. It also induces jumps from one branch to another, with the result that both branches may occur along either a downward or upward sweep of the dc bias. By comparing the red CVCs (those with the higher noise intensity) to the blue CVCs (those with the lower noise intensity),  we see that the jump frequency and position (the value of $i_{\rm dc}$ at which the jump occurs), also depends on the noise intensity. At the higher noise intensity the two branches are somewhat less developed, as one might expect. However, the important point to notice is that both branches are robust to the added noise at $i_{\textrm{dc}} \approx 1.44$, indicated by the vertical dashed lines in Figs.~\ref{fig9} (b) and (c).  For the case of the higher noise level shown in (b), both branches are present along the downward sweep in current, while for the case of the lower noise level shown in (c), both branches occur along the upward sweep.

\subsection{Switching in the presence of thermal fluctuations} \label{subsec:lifetime}
In Section~\ref{sec:Switching} we demonstrated that it is possible, without taking into account thermal fluctuations, to use the current pulse $i_p(t)$ to switch from the period-1 motion to a FP. We also saw that it was not possible to switch back from the fixed point to the period-1 motion. In this subsection, we will repeat our simulations of switching with the added thermal fluctuations.

\begin{figure}
    \centering
    \includegraphics[width=0.48\textwidth]{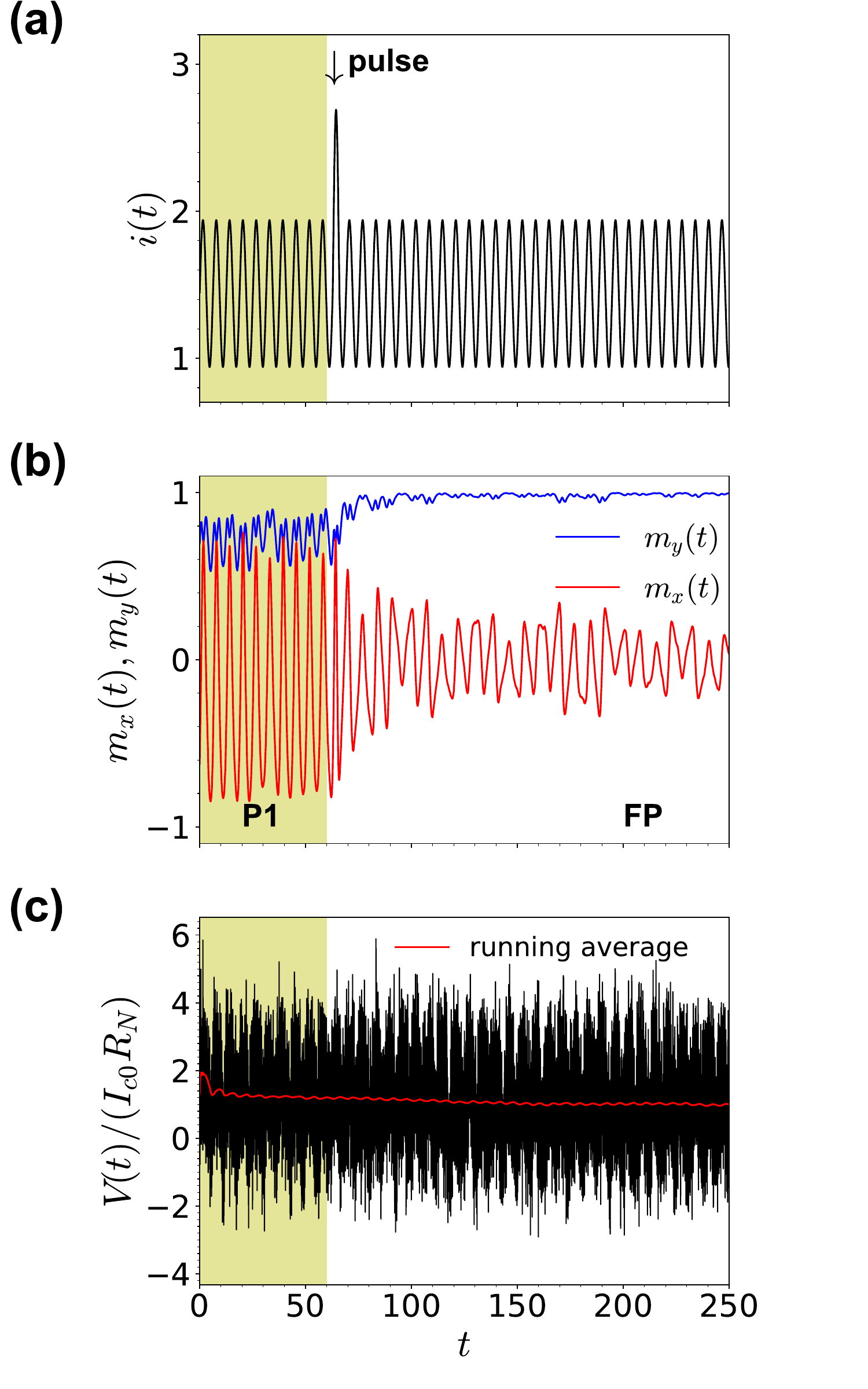}
    \caption{Magnetization dynamics in the type (ii) system under the current pulse in presence of the thermal fluctuations. (a) The applied current pulse is given by Eq.~(\ref{eq8}) and is shown by the black arrow. The parameters of the system and the current pulse are the same as in Figs.~\ref{fig2}~\ref{fig3} and~\ref{fig8}. (b) The switching from period-1 motion (P1 - shaded regions) with $V/(I_{c0}R_N) = 1.20$ to the fixed point (FP) $\mathbf{m} = (0,\ 1,\ 0)$ with $V/(I_{c0}R_N) = 0.95$. (c) Time dependence of the voltage along JJ is blurred by the thermal fluctuations. The running average (dashed red line) has been computed over 50 time units, using 5000 samples. The noise intensities $D_I = 5\times 10^{-4}$, $D_H = 4\times 10^{-5}$ are used for (b) and (c).}
    \label{fig10}
\end{figure}

In Fig.~\ref{fig10}, we show the same experiment as in Fig.~\ref{fig8}, performed here with the added thermal fluctuations. We choose the lower noise intensity level corresponding to the two voltage branches shown in Fig.~\ref{fig9} (c), at $i_{\textrm{dc}} = 1.44$. In the presence of these thermal fluctuations, we see that the switching from the period-1 motion to the FP is still possible.

\begin{figure}
    \centering
    \includegraphics[width=0.48\textwidth]{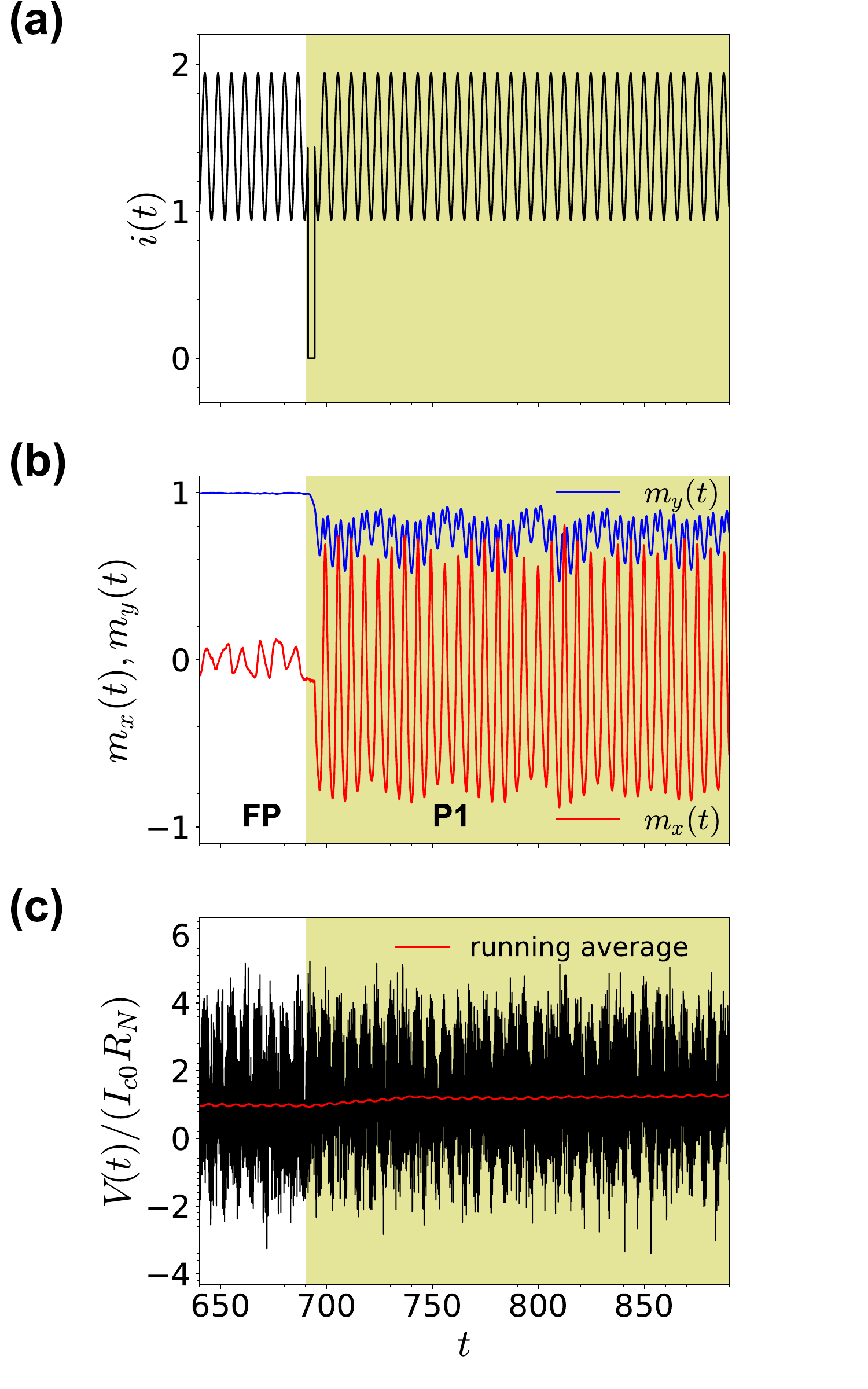}
    \caption{Magnetization dynamics in the type (ii) system under the short-time absence of the applied current in presence of the thermal fluctuations. (a) The applied current $i(t)$ is switched off for $\tau/2$ at $t_0 = 110\tau$. The parameters of the system are the same as in Figs.~\ref{fig2}~\ref{fig3}~\ref{fig8}, and~\ref{fig10}. (b) The switching from the fixed point (FP) $\mathbf{m} = (0,\ 1,\ 0)$ with $V/(I_{c0}R_N) = 0.95$ to the period-1 motion (P1 - shaded regions) with $V/(I_{c0}R_N) = 1.20$. (c) Time dependence of the voltage along JJ is blurred by the thermal fluctuations. The running average (dashed red line) has been computed over 50 time units, using 5000 samples. The noise intensities $D_I = 5\times 10^{-4}$, $D_H = 4\times 10^{-5}$ are used for (b) and (c).}
    \label{fig11}
\end{figure}

As mentioned in connection with Fig.~\ref{fig8}, in our simulations without the added thermal fluctuations, it appeared to be impossible to switch back from the FP, with $V/(I_{c0}R_N) = 0.95$, to the period-1 motion with $V/(I_{c0}R_N) = 1.20$. It was not possible because the torque $\mathbf{T}_2$ always acts parallel to $\mathbf{m}$ at the fixed points $(0, \pm 1,0)$. In the absence of thermal fluctuations, if one switches off the total current $i_{\textrm{dc}}$, the fixed points become unstable, but there is nothing in the simulation to perturb $\mathbf{m}$ away from one of the fixed points. The magnetization therefore remains on one of these points. In the presence of thermal fluctuations, however, this situation changes. We find that, after switching off the total current for a short time interval, the thermal fluctuations drive the system away from the unstable fixed point and, when the total current is switched back on, the magnetization may return to the period-1 motion. This transition can be induced reliably, provided that the time interval over which the total current is switched off is chosen so that the magnetization is perturbed from the FP into the basin of attraction for the period 1 motion. An example of this switching back, from the fixed point to the period-1 motion, is provided in Fig.~\ref{fig11}. In this case, we achieved the switching by simply turning off the total current for half a drive cycle.
Our simulations thus demonstrate that switching in presence of the added thermal fluctuations is indeed feasible. In fact, the instability caused by the added thermal fluctuations makes it possible to switch the system in both directions, as we have seen.

\section{Conclusion}
\label{sec:Conclusion}

We have studied the effect of the current-induced magnetization dynamics on the CVCs for three types of overdamped $\varphi_0$ Josephson junctions with dc and ac biases. We have shown that there exist three different scenarios for this effect depending on the system under consideration. In type (i), for the S/FM + Rashba/S structures, the magnetization dynamics has no effect on CVCs. In type (ii), for the S/FI/S on TI structures, the magnetization dynamics may lead to the presence of chaos and hysteresis with multiple branches in CVCs. In type (iii), for the S/FM/S on TI structures, the appearance of chaotic regimes and hysteresis is possible, even in the absence of the capacitance and without ac driving. Due to the additional magnetization degrees of freedom, even the overdamped $\varphi_0$ Josephson junction may exhibit complex dynamical regimes that are reflected in its current-voltage characteristics as hysteresis and chaos. Moreover, these dynamical regimes may exist even in absence of the ac driving. As the presence of the chaos and the hysteresis can affect potential uses of such systems in, for example, superconducting memory, we believe our results may add value to the emerging field of spintronics. We have also discussed the possibility of applying the hysteresis found in our simulations for the current pulse induced switching between the voltage states of the junctions under consideration, at a fixed dc-bias. Our simulations support the idea that switching is also possible in the overdamped limit, and even in the presence of thermal fluctuations. In fact, the instability provided by the added thermal fluctuations makes it possible to switch the type (ii) system back and forth between its two different voltage states, one being related to period-1 behavior of the magnetization dynamics, the other to a fixed point. Such switching properties are very promising for potential applications. They may serve as the source of voltage signals in rapid single flux quantum logic schemes, or as energy dependent memory, where the written information is encoded in the voltage of the junction.

Finally, from a dynamical systems point of view, the transition we found in the type (iii) system at $i_{\mathrm{dc}} = 1$, from the stable fixed point, immediately to chaos, is extremely interesting. Since there are few instances of such behaviour in the literature -- a notable exception being~\cite{avr16} -- it would be well worth further investigation. In future work we may try to establish, rigorously whether or not the type (iii) system may indeed present a fundamentally different, or at least rare, route to chaos.

\section{Acknowledgments}
We are thankful to Ya.~V.~Fominov for valuable discussions. A.~A.~M. acknowledges the financial support by the AYSS grant No. 23-302-02. The numerical calculations within this work were made possible with the help of the Russian Science Foundation (within the framework of project 22-42-04408) and the High Performance Computing facilities at the University of South Africa (Unisa). Yu. M. S. acknowledges financial support of Russian Science Foundation, project 22-71-10022 (analysis and discussion of the obtained numerical results). A. E. B. gratefully acknowledges CRC1 funding from the College of Science, Engineering and Technology at Unisa to participate in the 16th CHAOS 2023 International Conference. 

\section*{Appendix: Broken one-to-one corre- spondence between \texorpdfstring{$\mathbf{m}(t)$}{} periodicities and CVCs for the type (ii) system}
To understand why the one-to-one correspondence is broken, we notice that the function $\Phi = \varphi - r m_y$, in Eq.~(\ref{eq6b}), depends indirectly (and hence nonlinearly) on $m_x$, via the critical current $i_c(m_x)$. Thus, the frequency spectra of the time series $i_c(m_x(t))$ and $m_x(t)$ can contain very different components and we need to consider how time-dependent deviations of the critical current from its mean (rather than the oscillations of $m_x$), perturb the JJ. To this end, we expanded the perturbation, $\delta  i_c(t) = i_c(m_x) - \overline{i_c(m_x)}$, in a Fourier series with the frequencies $\Omega_i = \lambda_i \Omega$, where the $\lambda_i$ are positive numbers describing the harmonics ($\lambda_i = 1,2, 3, \ldots$) and subharmonics ($\lambda_i = 1/2, 1/3, \ldots$) of the ac drive frequency, $\Omega$.  For a fixed value of $i_{\textrm{dc}}$, the Josephson oscillations have the frequency $\omega_J = \overline{\dot{\Phi}}$. For frequency locking to occur, we must have $ \omega_J = k \Omega + \sum_j m_j \lambda_j \Omega $, where the term $k \Omega$ comes from the ac drive, and the terms $m_j \lambda_j \Omega$ come from the perturbation, $\delta  i_c(t)$. As we have seen in Fig.~\ref{fig2}, typically, for integer Shapiro steps, such frequency locking will occur when $k$ and $m_j \lambda_j$ are integers.
\begin{figure}[htb!]
    \centering
    \includegraphics[width=0.47\textwidth]{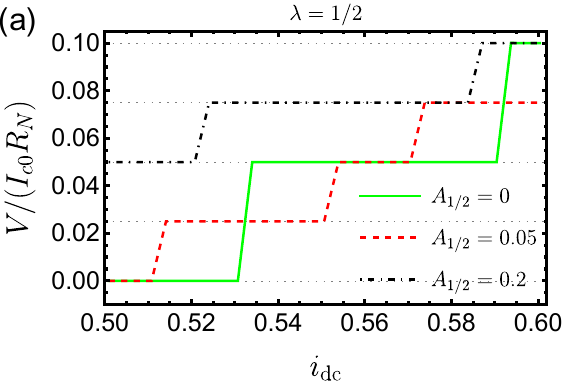}
    \includegraphics[width=0.47\textwidth]{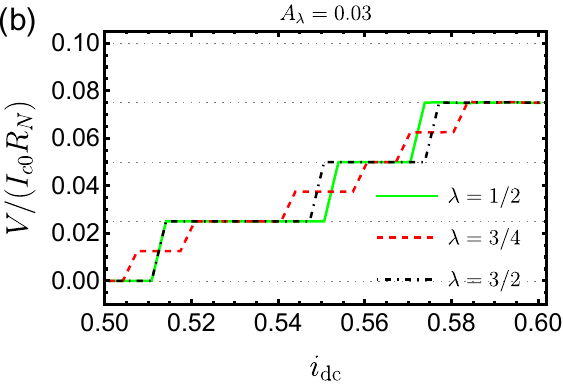}
    \caption{Stability of the first Shapiro step $\omega_J = \Omega$ against weak harmonic perturbations for changing the amplitudes of $A_{1/2}$ (a) and  $\lambda$ (b). The parameters used are $w = 0.05$, $\overline{i_{c}(m_x)}=0.9$, $A_1 = 0.1$, $i_{\textrm{ac}} = 0.5$, $\Omega = 1$.}
    \label{fig12}
\end{figure}

Now, let us focus on two specific ICs: $\mathbf{m}_1(0)$ and $\mathbf{m}_2(0)$, each of which leads to a different magnetization mode at the same value of $i_{\textrm{dc}}$. We assume that each of the magnetization modes, $\mathbf{m}_1(t)$ and $\mathbf{m}_2(t)$, corresponds to two different Shapiro steps, with the average voltages being $V_1$ for $\mathbf{m}_1(t)$ and $V_2$ for $\mathbf{m}_2(t)$. If the spectra of $\delta i_{c}(m_{x1})$ and $\delta i_{c}(m_{x2})$ contain the same subset of frequencies, $\lambda_j\Omega$, $j = 1,2,...$ -- which are responsible for the synchronization between the Josephson oscillations and the magnetic oscillations -- then we find that $V_1 = V_2$. Note that this equality of the voltages does not rely on the amplitudes of the various harmonics of $\delta i_{c}(m_{x1})$ and $\delta i_{c}(m_{x2})$ being the same. It depends only whether the two harmonic series have the same set of the frequencies $\lambda_j \Omega$, $j = 1,2,...\ $. Moreover, the spectra of $\delta i_{c}(m_{x1})$ and $\delta i_{c}(m_{x2})$ may contain small amplitude harmonics with mismatched frequencies, without affecting $V_1 = V_2$.

If the spectrum of $\delta i_{c}(m_{x1})$ contains at least the one term $A_{1k}\sin\left(\lambda_k\Omega t + \chi_{1k}\right)$ in which $A_{1k}$ significantly differs from the corresponding amplitude $A_{2k}$ of the spectrum for $\delta i_c(m_{x2})$, then one may expect to see $V_1 \neq V_2$. Typically, the bifurcations which are responsible for the birth of subharmonics of $\Omega$ in $\mathbf{m}(t)$ do not rearrange greatly the amplitudes of $\delta i_{c}(t)$. That is why we do not see the high periodicities $(\geq 6)$ of $\mathbf{m}(t)$ in the CVCs. 

In order to clarify these ideas via a concrete example, we replace the full dependence $i_{c}(m_x)$ with the simplified, approximate function: $\overline{i_c(m_x)} + A_{\lambda}\sin\left(\lambda \Omega t\right) + A_{1}\sin\left( \Omega t\right)$.
Specifically, we insert this approximate function into the RSJ model equation (\ref{eq6b}), and then recompute the Shapiro steps for different values of $A_{\lambda}$ and $\lambda$. The results of these calculations are presented in Fig.~\ref{fig12}.  
We see in Fig.~\ref{fig12}(a) that there exist an interval, $0.554\leq i_{\textrm{dc}} \leq 0.570$, for the first Shapiro step, $\omega_J = \Omega$, for which $A_{1/2} = 0.05$ produces the same voltage as  $A_{1/2} = 0$. Also, $A_{1/2} = 0.2$ is so large that it shifts the current interval for the first Shapiro step to the left side of the picture. Similarly, in Fig.~\ref{fig12}(b) we find that none of the subharmonics, $A_{1/2}\sin\left(\Omega t/2\right)$, $A_{3/4}\sin\left(3 \Omega t/4\right)$, or $A_{3/2}\sin\left(3\Omega t/2\right)$, shift the voltage of the first Shapiro step in the current interval $0.561\leq i_{\textrm{dc}}\leq 0.567$.  Therefore, for small enough harmonic perturbations the voltages of the two Shapiro steps remain the same, besides the voltages of the additional steps due to the spectral form of the perturbation.

Close inspection of the CVC shown in Fig.~\ref{fig2}(b), in the region $i_{\textrm{dc}} \approx 1$, also reveals half-integer Shapiro steps. On these half-integer steps we found that the spectrum of $\delta i_{c}(m_x)$ contains subhamonics of $\Omega$ which are approximately of the same order of magnitude as the harmonics of $\Omega$. Thus, the same locking mechanism appears to apply to both types of steps.

\bibliography{mazanik.bib}
\end{document}